\documentclass[prb,twocolumn,showpacs,preprintnumbers]{revtex4-2}

\usepackage{amsmath, amssymb, graphicx, hyperref, subcaption}
\usepackage[labelfont=bf]{caption}
\usepackage[normalem]{ulem} 
\usepackage{placeins}       
\hypersetup{colorlinks=true, linkcolor=blue, citecolor=blue, urlcolor=blue}

\begin{document}

\title{
Shot noise as a probe for Andreev reflection in graphene-based heterojunctions}

\author{Shahrukh Salim and Poornima Shakya}
\affiliation{Department of Physics, KIAS, Seoul, South Korea}
\affiliation{Department of Physics, IISER Bhopal, India}

\begin{abstract}
Shot noise emerges due to the discrete nature of charge transport and provides direct access to the underlying microscopic transport mechanisms governing current flow in mesoscopic conductors. In this work, we demonstrate that quantum shot noise offers a direct and robust fingerprint of Andreev reflection (AR), distinguishing between retro (RAR) and specular (SAR) processes in graphene–superconductor (GS), graphene–superconductor–graphene (GSG), and superconductor–graphene–superconductor (SGS) junctions.
At the GS interface, exact reflection amplitudes obtained from full wavefunction matching within the Bogoliubov-de Gennes formalism capture RAR and SAR regimes. The associated Fano factor exhibits distinct Fermi-level-dependent signatures, with RAR suppressing and SAR enhancing the shot noise. 
Extensions to GSG and SGS configurations reveal how the transmission spectrum and, consequently, the noise profile are modified in the presence of multiple interfaces, coherent quasiparticle interference, and superconducting phase variations. Our findings establish shot noise spectroscopy as a potent and experimentally viable probe for differentiating Andreev reflection types in graphene-based quantum devices, providing complementary insights beyond conventional conductance measurements.
\end{abstract}

\maketitle

\section{Introduction}\label{sec:intro}
Shot noise originates from the quantization of electric charge and serves as a powerful probe of the fundamental microscopic processes that dictate electron transport in mesoscopic systems
\cite{BlanterButtiker2000,LevitovLeeLesovik1996,Nazarov2003Book,Belzig1999Review}. While conductance measurements yield averaged information about transmission, shot noise provides direct access to current–current correlations \cite{Buttiker1992,MartinLandauer1992} and thus reveals information about quantum coherence \cite{Oberholzer2001AB,Henny1999}, partition statistics \cite{Lesovik1989,Reznikov1995}, and the interplay between different transport channels \cite{BeenakkerButtiker1992}. This makes noise spectroscopy valuable, particularly, in systems where coherent electron–hole processes dominate the subgap transport, such as Andreev reflection \cite{Andreev1964,BTK1982,Beenakker1992NSreview,deGennes1999} and crossed Andreev conversion in multiterminal superconducting devices \cite{Falci2001,Beckmann2004,Russo2005}. In composite devices that couple graphene to a superconductor, shot noise is profoundly influenced by the interplay between graphene’s relativistic-like quasiparticles and the induced superconducting correlations at the interface \cite{Beenakker2006SpecAR,Beenakker2008GrapheneColloquium,Heersche2007,Calado2015,Mizuno2013,BenShalom2016}.

Graphene offers a unique platform to investigate such physics due to its gapless, linear dispersion and massless Dirac fermions whose transport properties can be continuously tuned via electrostatic gating \cite{CastroNeto2009,Beenakker2006Klein}. When a graphene sheet is placed in contact with a superconductor, Cooper pair correlations penetrate into the graphene through the proximity effect, enabling Andreev reflection (AR) at the interface \cite{TinkhamBook1996,Andreev1964,deGennes1999}. In the conventional retro Andreev reflection (RAR) process, an incoming electron with energy below the superconducting gap is reflected as a hole retracing the incoming trajectory, while a Cooper pair is transmitted into the superconductor \cite{BTK1982,Beenakker1992NSreview,Zaitsev1984,KupriyanovLukichev1988}. The relativistic band structure of graphene, however, permits an additional inter-band process - specular Andreev reflection (SAR) - where the reflected hole resides in the opposite energy band, leading to a specular rather than retrograde trajectory \cite{Beenakker2006SpecAR,Nilsson2007,Linder2008,BlackSchaffer2008,Efetov2015}. SAR emerges near the charge neutrality point when the Fermi energy is comparable or smaller than the excitation energy, a phenomenon absent in conventional metal–superconductor junctions \cite{Beenakker2006SpecAR,Katsnelson2006}.

From the standpoint of current fluctuations, Andreev processes modify the effective transferred charge and alter the distribution of transmission probabilities across available channels \cite{Khlus1987,AnantramDatta1996,BlanterButtiker2000,Belzig1999Review}.
In the subgap regime, perfect AR doubles the transferred charge to $2e$, potentially enhancing the noise compared to the normal state \cite{Khlus1987}, whereas when interface transparency is high and AR dominates, the nearly ballistic channels suppress fluctuations and reduce the Fano factor below the Poissonian limit \cite{deJongBeenakker1994,BlanterButtiker2000,Nazarov1999Circuit}.
Near the Dirac point, SAR becomes relevant resulting in inter-band kinematics and angular selectivity reducing the effective transparency for many modes, thus, increasing partition noise and imprinting distinct gate- and bias-dependent features on the noise spectrum \cite{Beenakker2006SpecAR,DiCarlo2008,Danneau2008,BenShalom2016,Calado2015}. 
This sensitivity of shot noise to the relative weight of RAR and SAR provides a direct diagnostic of the dominant Andreev process and of interface quality \cite{Beenakker2008GrapheneColloquium}.

In this work, we develop a unified framework for quantum shot noise in graphene-based superconducting hybrid junctions, enabling a direct connection between microscopic scattering properties and measurable shot noise characteristics \cite{TinkhamBook1996,Beenakker1992NSreview,deGennes1999}. Beginning with the graphene–superconductor (GS) interface as the minimal geometry, exact solutions of the Bogoliubov–de Gennes (BdG) equations provide analytic form of the normal and Andreev reflection amplitudes whose dependence on quasiparticle incidence angle, excitation energy, Fermi level, and interface transparency directly determines the noise response \cite{TinkhamBook1996,Beenakker1992NSreview,BTK1982,deGennes1999,Belzig1999Review}. In the RAR regime, nearly perfect Andreev conversion at transparent contacts suppresses noise \cite{deJongBeenakker1994}, whereas in the SAR regime, angular dependence of the interband matching constraints modulate Andreev probability and enhance fluctuations, often yielding oscillatory gate- and bias-dependent features due to the presence of interference effects \cite{Beenakker2006SpecAR,Cayssol2008,Rickhaus2015,TanakaKashiwaya2000,Zaitsev1984,KupriyanovLukichev1988}. Extending the analysis to graphene–superconductor–graphene (GSG) junctions, the finite superconducting slab forms a resonant cavity for electron–hole excitations, producing oscillatory gate-tunable modulations in the Fano factor \cite{CuevasBelzig2003,NazarovBlanterBook2009,Rickhaus2015}, while in superconductor–graphene–superconductor (SGS) Josephson junctions, Andreev bound states (ABS) facilitate phase-coherent transport and multiple Andreev reflection (MAR) processes under finite bias generate subharmonic features in the differential noise and bias dependence of the Fano factor \cite{Golubov2004,TitovBeenakker2006,Likharev1979,Bretheau2013,OctavioTinkham1983,AverinBardas1995,BardasAverin1997,Bratus1995,Cuevas1996,Cuevas1999Noise}. The interplay of RAR and SAR modifies ABS dispersions, leading to qualitative changes in phase-dependent shot noise \cite{Nilsson2007,Golubov2004,TitovBeenakker2006}. Our results demonstrate that shot noise spectroscopy serves not merely as a supplementary measurement but as a primary diagnostic for identifying Andreev processes, assessing interface quality, and exploring phase-coherent transport in graphene-based quantum devices \cite{BlanterButtiker2000,Beenakker2006SpecAR,Efetov2015}.

The manuscript is structured as follows:
Section~\ref{sec:gs} describes the Hamiltonian for the graphene–superconductor (GS) model and briefly explains the Bogoliubov-de Gennes formalism used to understand the retro and specular Andreev processes for GS, GSG and SGS junctions.
Section~\ref{sec:shotnoise} introduces the scattering-matrix framework for shot noise and the Fano factor calculated throughout the work.
Section~\ref{sec:results} presents our results for GS, GSG, and SGS junctions, including angle- and energy-resolved noise maps, Fano-factor trends, and temperature dependence of the shot noise.
Lastly, we summarise our results in Section~\ref{sec:conclusion} and provide an experimental outlook. Additional derivations related to the wavefunction matching, limiting forms, and amplitudes are provided in Appendices~\ref{appendix_wavefuncmatch} and \ref{appendix_retrospec}.

\section{Graphene-Superconductor (GS) based Junctions}\label{sec:gs}
In this section, we start from the graphene–superconductor (GS) junction, the minimal geometry that captures the essential physics of Andreev reflection and serves as the foundation for more complex hybrid structures \cite{Beenakker2008GrapheneColloquium,Heersche2007,Beenakker2006SpecAR,TitovBeenakker2006,Calado2015}. Using the transfer matrix formalism, we obtain analytical expressions for reflection and transmission amplitudes in both RAR and SAR regimes, incorporating the energy, angle, and doping dependence of quasiparticle scattering \cite{Beenakker2006SpecAR,Nilsson2007}. These results provide the basis for calculating shot noise and the Fano factor, thereby identifying distinct noise signatures of each reflection process \cite{BlanterButtiker2000,DiCarlo2008,Danneau2008}. The GS analysis also functions as a building block for GSG and SGS junctions, where additional interfaces introduce resonant interference and phase-coherent transport; although closed-form solutions are often inaccessible, extensions of the GS formalism yield symbolic scattering amplitudes that can be directly connected to experimentally measurable noise characteristics \cite{CuevasBelzig2003,Golubov2004,Nazarov1999Circuit,Efetov2015,BenShalom2016}.

\subsection{GS interface}

The low-energy excitations in monolayer graphene occur in the vicinity of two inequivalent Dirac points $K$ and $K'$ in the Brillouin zone. In the absence of interactions, the quasiparticle dynamics near each valley are governed by the massless Dirac equation. Restricting to a single valley and ignoring intervalley scattering (which is strongly suppressed for smooth interfaces), the effective Hamiltonian for the system takes the following form:
\begin{equation}\label{eq_Ham}
    H_G = \hbar v_F \left( \sigma_x k_x + \sigma_y k_y \right) - \mu \, \sigma_0 ,
\end{equation}
where $\sigma_{x,y}$ are Pauli matrices acting in the sublattice (pseudospin) space, $\sigma_0$ is the $2\times 2$ identity matrix, $v_F \approx 10^6~\mathrm{m/s}$ is the Fermi velocity, and $\mu$ denotes the chemical potential measured from the charge neutrality (Dirac) point. This Hamiltonian yields a linear energy dispersion $E_{\pm}(\mathbf{k}) = \pm \hbar v_F |\mathbf{k}| - \mu$, where the $+$ and $-$ branches correspond to conduction and valence band states, respectively. The pseudospin structure ensures that the carrier momentum and sublattice degree of freedom are locked, a feature that plays a key role in the unconventional Andreev reflection properties of graphene.
To account for proximity-induced superconductivity, we employ the Bogoliubov-de Gennes (BdG) formalism, which treats electron-like and hole-like excitations on an equal footing by embedding $H_G$ into a particle-hole basis. The resulting $4\times 4$ BdG Hamiltonian for the graphene-superconductor interface takes the form
\begin{equation}
    \mathcal{H}_{\mathrm{BdG}} =
    \begin{pmatrix}
        H_G & \Delta(x) \\
        \Delta^*(x) & -H_G^*
    \end{pmatrix},
\end{equation}
where $\Delta(x)$ is the complex-valued and spatially dependent superconducting pair potential, and $H_G^*$ is the complex conjugate of the graphene Hamiltonian acting in the hole sector. The diagonal blocks describe the normal-state dynamics of electrons and holes, while the off-diagonal blocks couple the two sectors via the superconducting condensate. This coupling is responsible for Andreev reflection: an incoming electron from the normal side can be converted into a reflected hole while a Cooper pair is transferred to the superconducting region.

In our model, we adopt a step-like profile for the Cooper pair potential,
\begin{equation}
    \Delta(x) =
    \begin{cases}
        \Delta_0 e^{i\phi}, & x < 0 \quad \text{(superconductor)}, \\
        0, & x > 0 \quad \text{(graphene)},
    \end{cases}
\end{equation}
which idealizes the GS interface as atomically sharp and free of disorder. Here, $\Delta_0$ is the magnitude of the superconducting gap, and $\phi$ is the macroscopic superconducting phase. This simple model captures the essential physics of proximity-induced pairing and allows us to derive analytical expressions for the scattering amplitudes. For $\mu \gg E$, the reflected hole lies in the same band as the incident electron, yielding RAR. In contrast, when $\mu \ll E$, the reflected hole occupies the opposite band, giving rise to SAR. This dependence is encoded in the matching conditions at the GS interface, where the pseudospin structure of the Dirac quasiparticles and the superconducting coherence factors together determine the reflection and transmission amplitudes.

\subsection{GSG Junction}

The graphene–superconductor–graphene (GSG) junction consists of a superconducting segment of finite width $d$ embedded between two semi-infinite graphene leads. This configuration serves as a natural extension of the single-interface GS system, introducing two interfaces at $x = 0$ and $x = d$ where quasiparticles can undergo multiple normal and Andreev reflections. The resulting interplay between electron–hole conversion at each interface and phase accumulation within the superconducting region gives rise to rich interference phenomena, including Fabry–P\'erot–like resonances in the subgap regime \cite{Cayssol2008,Rickhaus2015,CuevasBelzig2003}.

To analyze transport in this geometry, we solve the BdG equations in each region and impose continuity of the four-component spinor wavefunctions at both interfaces. The matching conditions at $x = 0$ connect the incident and reflected amplitudes in the left graphene lead to the quasiparticle amplitudes inside the superconducting slab, while the matching condition at $x = d$ connects the latter to the transmitted and reflected amplitudes in the right graphene lead. These two sets of interface conditions can be compactly expressed using the transfer matrix formalism, leading to a composite transfer matrix $M_{\mathrm{GSG}}$ that relates the electron and hole amplitudes on the left and right sides of the junction \cite{NazarovBlanterBook2009,Beenakker1992NSreview}.
\begin{equation}
\begin{pmatrix}
\mathbf{A}_\mathrm{R} \\
\mathbf{B}_\mathrm{R}
\end{pmatrix}
=
M_{\mathrm{GSG}}
\begin{pmatrix}
\mathbf{A}_\mathrm{L} \\
\mathbf{B}_\mathrm{L}
\end{pmatrix},
\end{equation}
where $\mathbf{A}$ and $\mathbf{B}$ denote the electron and hole amplitude vectors, respectively, and the subscripts $\mathrm{L}$ and $\mathrm{R}$ refer to the left and right graphene leads.

The structure of $M_{\mathrm{GSG}}$ encodes all scattering processes in the junction, including normal transmission, normal reflection, local Andreev reflection (LAR), and crossed Andreev reflection (CAR). By decomposing this matrix, one can extract the corresponding reflection and transmission coefficients, which determine the conductance, shot noise, and other transport observables \cite{AnantramDatta1996,BlanterButtiker2000}. The explicit form of $M_{\mathrm{GSG}}$, together with the  individual scattering amplitudes, appears in one of our earlier works \cite{salim2024effect}.

\subsection{SGS Junction}

The superconductor–graphene–superconductor (SGS) junction is the natural counterpart of the GSG configuration, obtained by reversing the geometry, so that a graphene segment of finite length $L$ is embedded between two semi-infinite superconducting electrodes. This setup is of particular interest because the finite graphene region acts as a ballistic or quasi-ballistic weak link, and transport is mediated by phase-coherent Andreev bound states (ABS) formed through multiple electron–hole conversions at the two GS interfaces \cite{TitovBeenakker2006, Golubov2004}.

A key control parameter in the SGS junction is the superconducting phase difference defined as 
\begin{equation}
\phi = \phi_1 - \phi_2,
\end{equation}
where $\phi_{1,2}$ are the macroscopic phases of the left and right superconducting contacts, respectively. This phase difference directly modulates the ABS spectrum and, consequently, the supercurrent through the junction. Within the transfer-matrix framework, matching the four-component spinor wavefunctions at $x = 0$ and $x = L$ yields a total transfer matrix $M_{\mathrm{SGS}}$ that connects quasiparticle amplitudes across the graphene weak link. The bound-state condition can then be written as
\begin{equation}
\big| M_{\mathrm{SGS}} \big| = e^{\pm 2i\phi},
\end{equation}
which explicitly reflects the phase dependence of the quantized ABS energies \cite{Golubov2004}.

The structure of $M_{\mathrm{SGS}}$ contains all relevant scattering processes - normal reflection, local Andreev reflection, and crossed Andreev reflection - occurring at each interface, as well as the phase accumulation due to propagation through the graphene segment. Solving the bound-state equation provides the ABS dispersion $E(\phi)$, from which one can compute the Josephson current, temperature dependence, and sensitivity to gate voltage and doping level. The same framework also enables calculating finite-bias transport properties, such as multiple Andreev reflection (MAR) features in the differential conductance and associated shot noise. The explicit form of $M_{\mathrm{SGS}}$ and the derivation of the corresponding resonance and transmission coefficients appear in one of our earlier works \cite{salim2023revisiting}. 

In the following section, we introduce the concepts of quantum shot noise and the Fano factor, which constitute the central findings of our paper.

\section{Shot Noise in Graphene-Based Junctions}\label{sec:shotnoise}

Shot noise refers to the temporal fluctuations of electric current that arise from the discrete nature of charge carriers \cite{BlanterButtiker2000,NazarovBlanterBook2009}. In mesoscopic conductors, where quantum coherence and phase-sensitive scattering play a central role, shot noise provides information that is not accessible from conductance measurements alone \cite{LevitovLeeLesovik1996,Belzig1999Review}. It probes the statistical correlations between transmission events, revealing the interplay between quantum interference, carrier partitioning, and interaction effects \cite{BlanterButtiker2000,NazarovBlanterBook2009}. In graphene-based Josephson junctions, shot noise is strongly influenced by Andreev reflection processes at the superconductor-graphene interfaces \cite{Beenakker2008GrapheneColloquium,TitovBeenakker2006,Calado2015}.

At a graphene–superconductor interface, Andreev processes fundamentally reshape current fluctuations by altering both the effective transferred charge and the distribution of transmission eigenvalues \cite{Khlus1987,AnantramDatta1996,BlanterButtiker2000}. The resulting shot noise depends sensitively on the competition between normal transmission, local Andreev reflection, and crossed processes, as well as on the angular and energy dependence imposed by graphene’s Dirac spectrum \cite{AnantramDatta1996,NazarovBlanterBook2009,Belzig1999Review}. When the Fermi level lies well above the excitation energy, intraband RAR dominates, and nearly transparent channels suppress fluctuations, reducing the Fano factor below its Poissonian value \cite{deJongBeenakker1994,BlanterButtiker2000}. Close to charge neutrality point, however, interband specular reflection (SAR) becomes prominent \cite{Beenakker2006SpecAR,Nilsson2007}, where angular selectivity lowers effective transparency and enhances partition noise and altering the bias and gate-voltage dependence of the Fano factor \cite{DiCarlo2008,Danneau2008,Calado2015}.

At zero temperature, the low-frequency shot noise power $S$ can be expressed within the scattering-matrix approach as \cite{BlanterButtiker2000,NazarovBlanterBook2009}
\begin{equation}
S = 2 e I F,
\qquad \text{with }
F = \frac{\sum_n T_n \left(1 - T_n\right)}{\sum_n T_n},
\end{equation}
where $I$ is the direct current and $T_n$ represents the transmission probability of the $n$-th transport channel. The term $T_n(1-T_n)$ reflects the fluctuation weight of each channel: perfectly transmitting ($T_n=1$) or completely closed ($T_n=0$) channels do not generate shot noise, whereas partial transmission ($0<T_n<1$) yields the strongest contribution \cite{BlanterButtiker2000,LevitovLeeLesovik1996}. In the presence of Andreev processes the charge transferred per event is modified, effectively replacing $e$ by $e^\ast \simeq 2e$ in the subgap regime, which enhances the overall noise amplitude \cite{Khlus1987,AnantramDatta1996,Beenakker1992NSreview,deJongBeenakker1994}.
This framework provides a direct correspondence between the microscopic scattering amplitudes derived from the BdG approach and the experimentally
measurable noise characteristics \cite{Beenakker1992NSreview,NazarovBlanterBook2009}. In the following sections, we employ it to demonstrate how RAR and SAR generate distinct shot-noise signatures in graphene-based hybrids, starting from the analytically solvable GS interface and then extending to GSG and SGS geometries, where multiple reflections and phase coherence crucially shape the shot noise response \cite{Beenakker2006SpecAR,Beenakker2008GrapheneColloquium,TitovBeenakker2006,Calado2015}.

\begin{figure}[t]
  \captionsetup{justification=raggedright,singlelinecheck=false}
  \centering
  \subfloat[]{%
    \includegraphics[width=0.48\columnwidth]{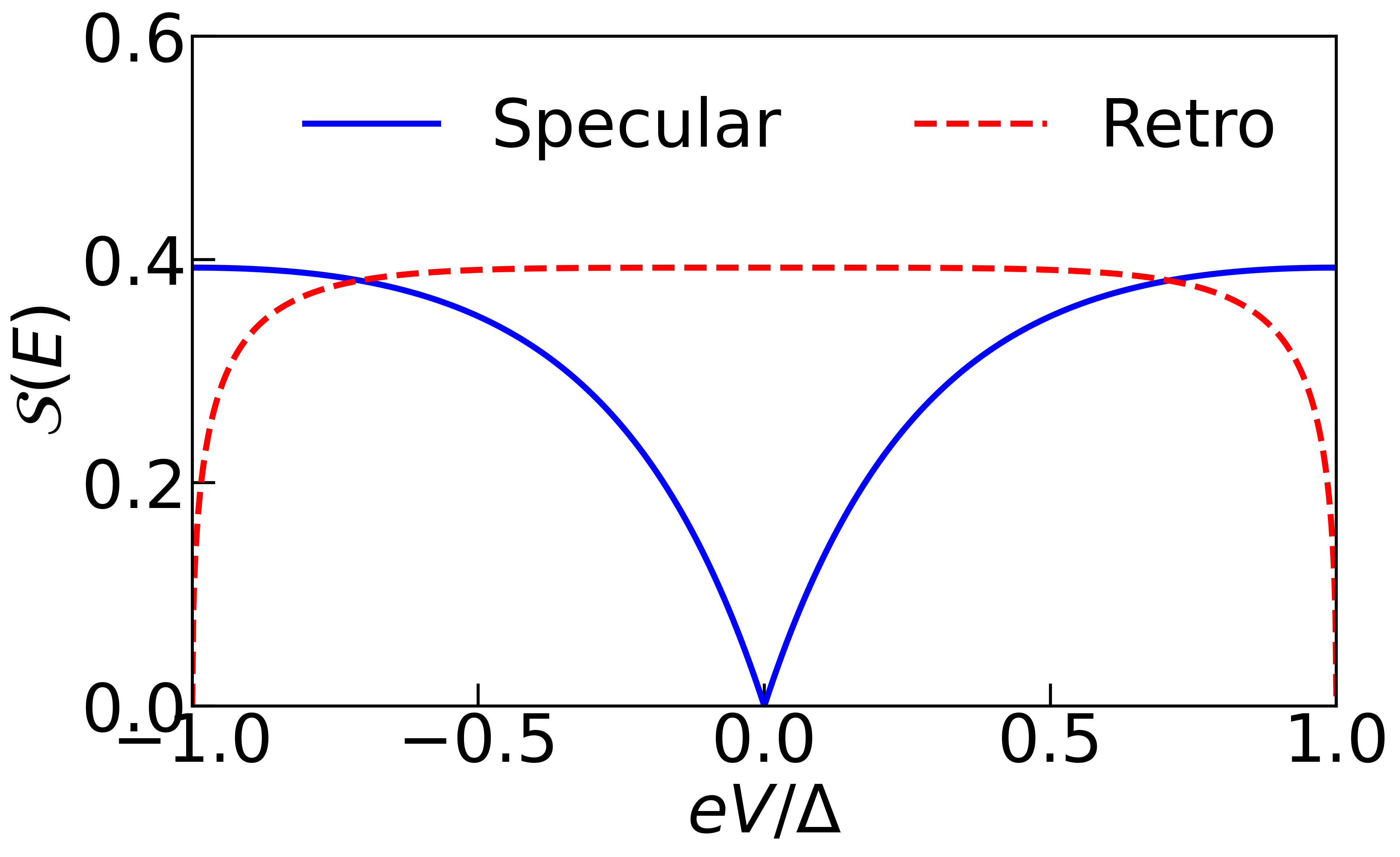}%
    \label{fig:shot_noise1a}
  }\hfill
  \subfloat[]{%
    \includegraphics[width=0.48\columnwidth]{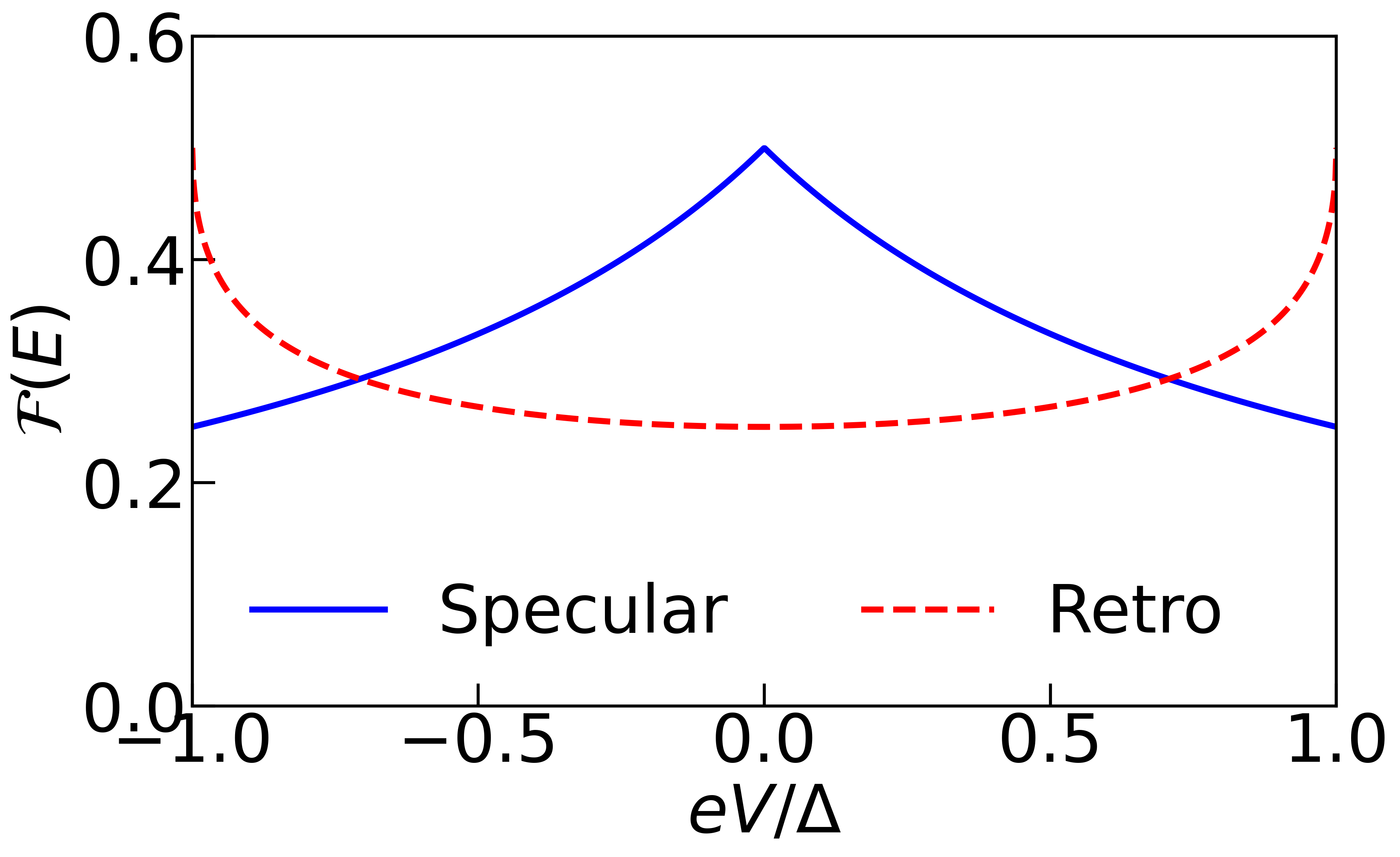}%
    \label{fig:shot_noise1b}
  }
  \caption{
  Angle-averaged subgap (a) shot noise \(S(E)\) and Fano factor \(F(E)\)
  for specular (blue, solid) and retro (red, dashed) Andreev reflection at a graphene–superconductor interface.
  The curves highlight the suppression near \(\lvert E\rvert\!\to\!\Delta\) in the retro case and the zero-bias minimum for the specular case.}
  \label{fig:shot_noise1}
\end{figure}
\begin{figure}[t]
      \captionsetup{justification=raggedright,singlelinecheck=false}
      \centering
    
      \begin{subfigure}[t]{0.32\textwidth}
        \centering
        \includegraphics[width=\linewidth]{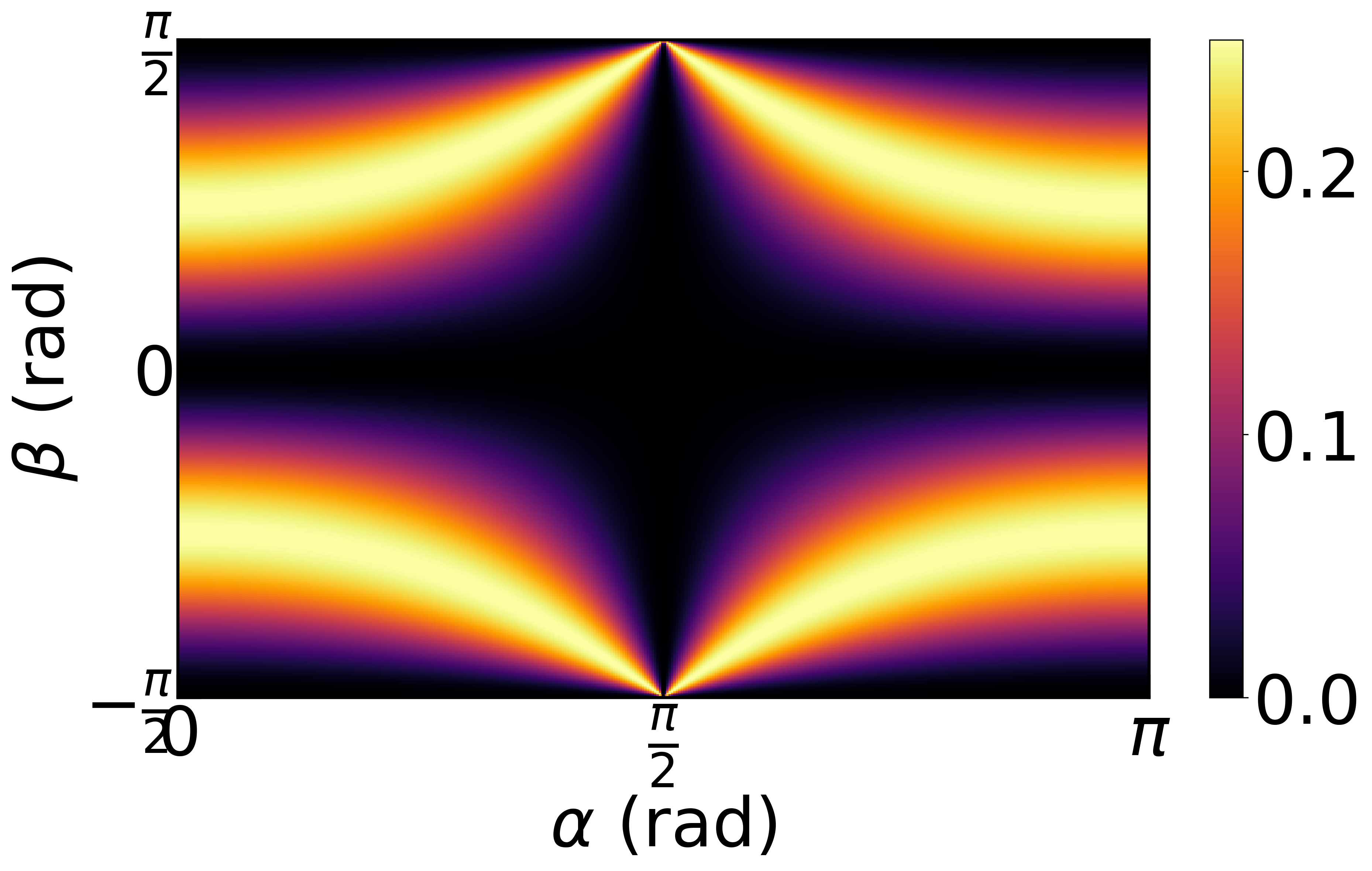}
        \caption{}
        \label{fig:shot_noise_combo:a}
      \end{subfigure}\hfill
      \begin{subfigure}[t]{0.32\textwidth}
        \centering
        \includegraphics[width=\linewidth]{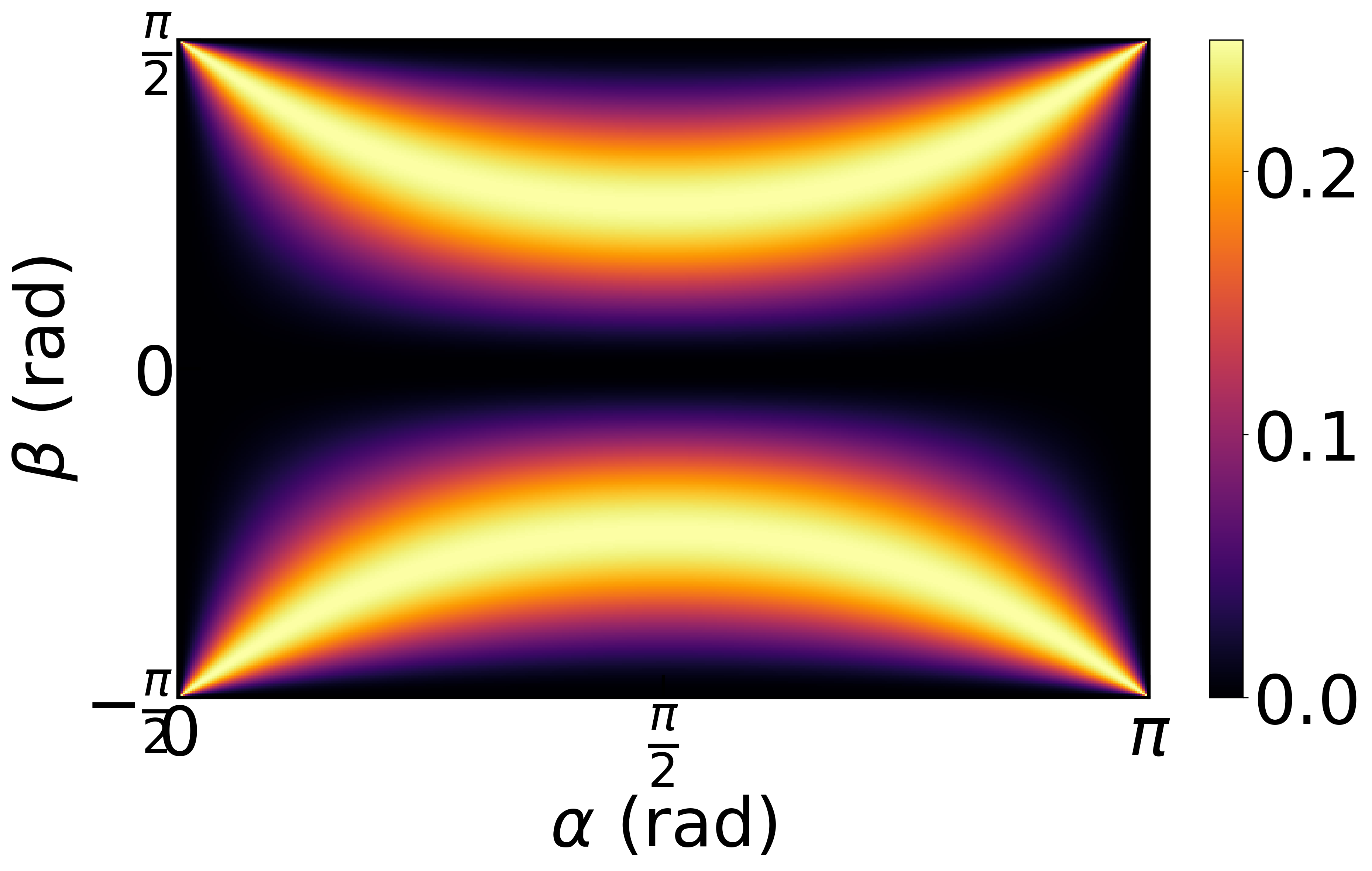}
        \caption{}
        \label{fig:shot_noise_combo:b}
      \end{subfigure}\hfill
      \begin{subfigure}[t]{0.32\textwidth}
        \centering
        \includegraphics[width=\linewidth]{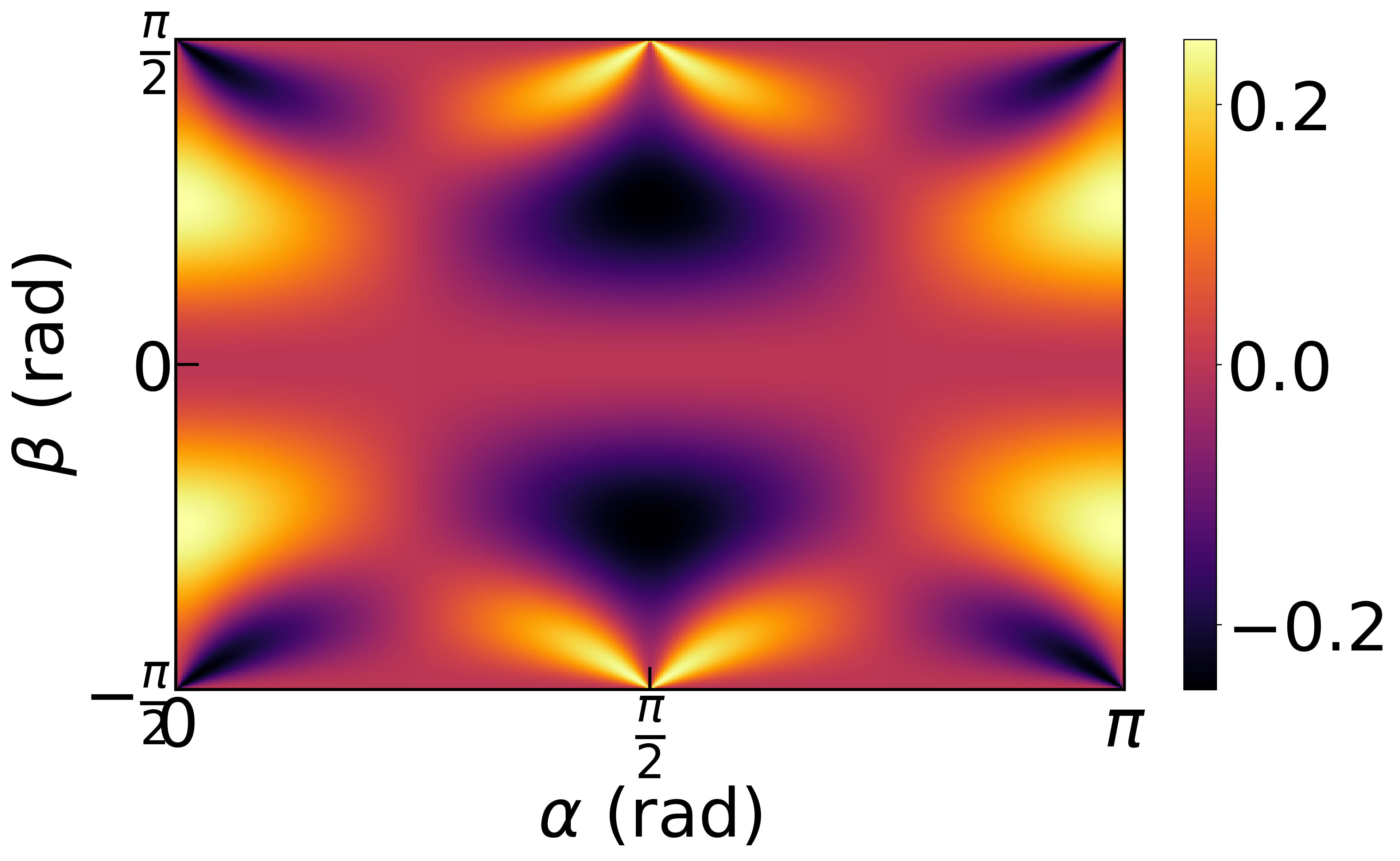}
        \caption{}
        \label{fig:shot_noise_combo:c}
  \end{subfigure}
  \caption{Angle–resolved shot–noise maps for a GS interface as a function of $\alpha$ (in radians) and $\beta$ (in radians):
  (a) specular \(S_{\mathrm{spec}}(\alpha,\beta)\), (b) retro \(S_{\mathrm{retro}}(\alpha,\beta)\),
  and (c) difference \(\Delta S(\alpha,\beta)=S_{\mathrm{spec}}-S_{\mathrm{retro}}\).
  Here, \(\alpha\in[-\pi/2,\pi/2]\) is the incidence angle and \(\beta=\arccos(E/\Delta)\in[0,\pi]\) encodes energy.}
  \label{fig:shot_noise_combo}
\end{figure}
\begin{figure}[h]
  \centering
  \begin{subfigure}[t]{0.32\textwidth}
    \centering
    \includegraphics[width=\linewidth]{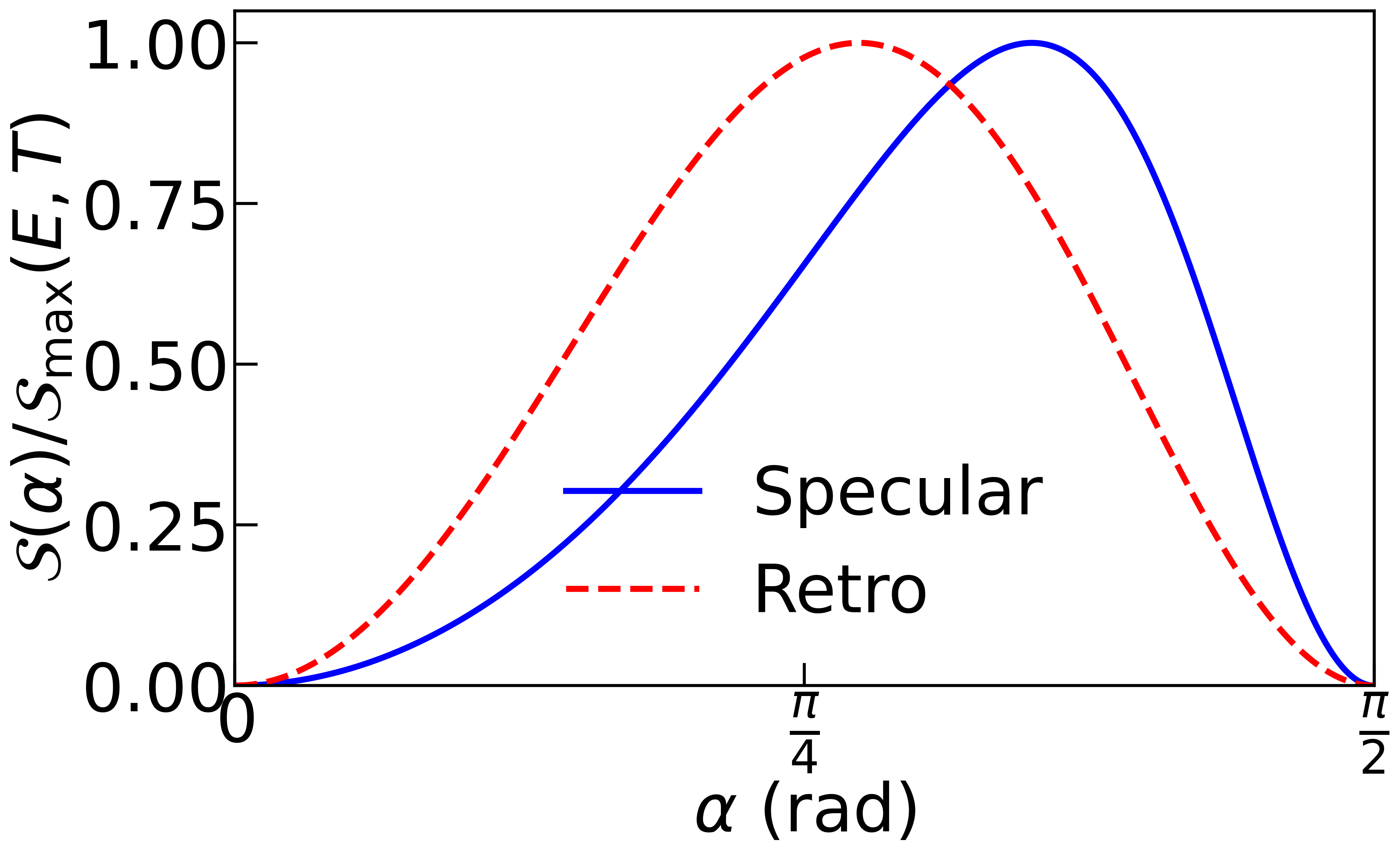}
    \caption{}
    \label{fig:sn_angle_T02}
  \end{subfigure}\hfill
  \begin{subfigure}[t]{0.32\textwidth}
    \centering
    \includegraphics[width=\linewidth]{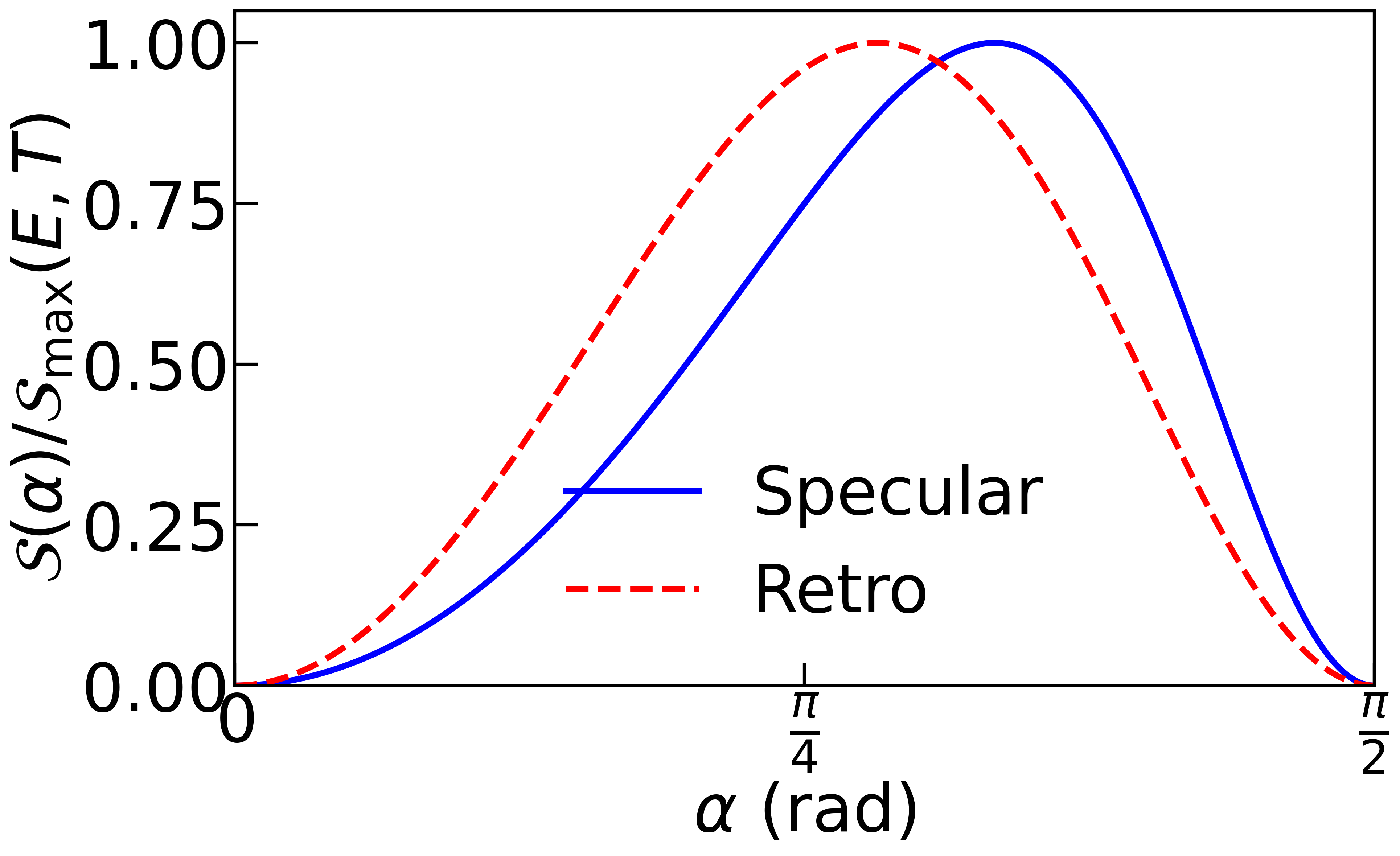}
    \caption{}
    \label{fig:sn_angle_T05}
  \end{subfigure}\hfill
  \begin{subfigure}[t]{0.32\textwidth}
    \centering
    \includegraphics[width=\linewidth]{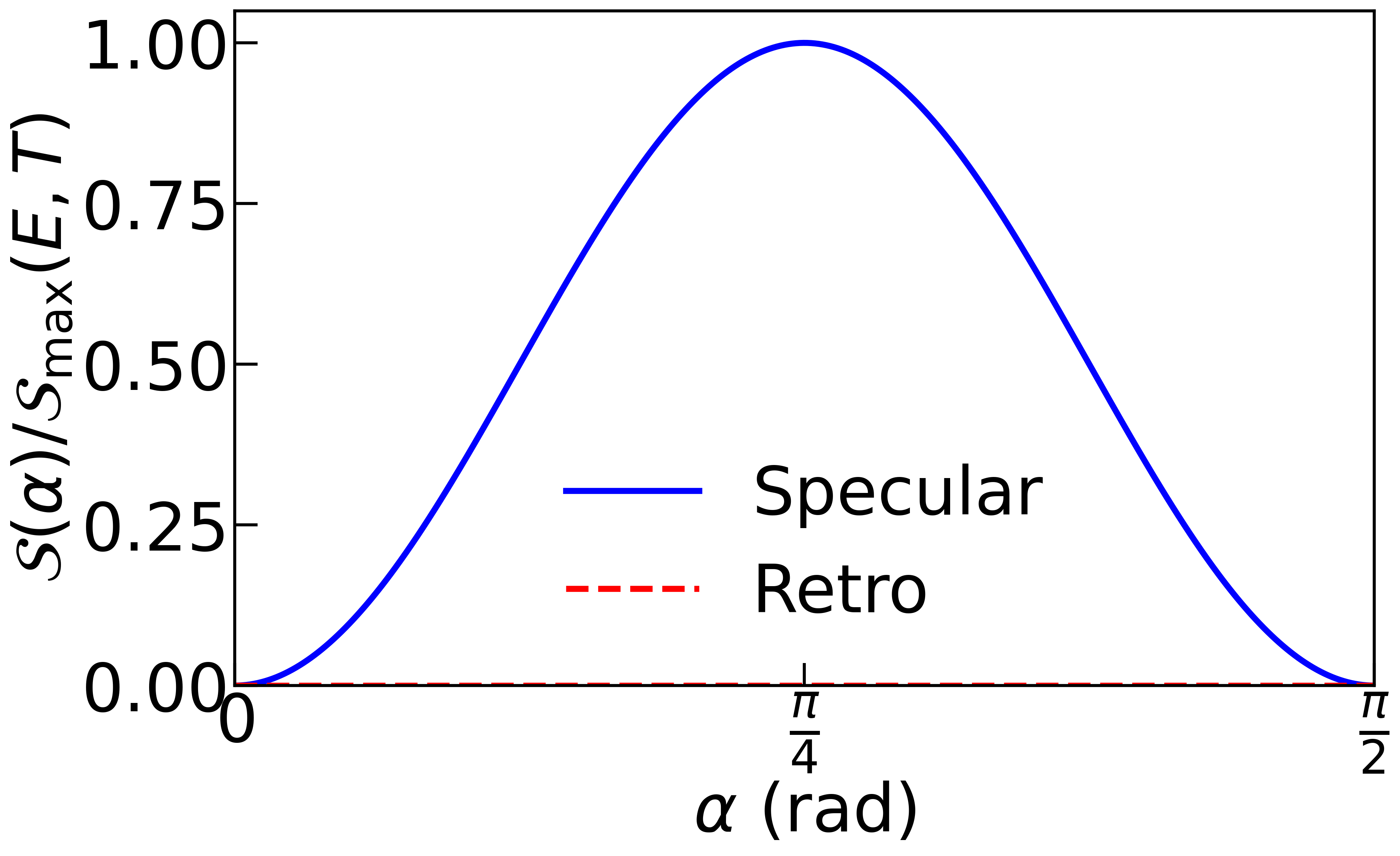}
    \caption{}
    \label{fig:sn_angle_T09}
  \end{subfigure}
  \caption{Normalized angular shot noise $S(\alpha)/S_{\max}(E,T)$ as a function of $\alpha$  (in radians) at different temperatures (a) $T=0.2T_c$, (b) $T=0.5T_c$ and (c) $T=0.9T_c$ for specular (blue, solid) and retro (red, dashed) Andreev processes.}
  \label{fig:sn_angle_norm_three}
\end{figure}

\section{Results and Discussions}\label{sec:results}

In this section, we present the calculated shot noise characteristics for the three device geometries considered: the single-interface graphene-superconductor (GS) junction, the graphene-superconductor-graphene (GSG) junction, and the superconductor-graphene-superconductor (SGS) Josephson junction. 
For each case, the starting point is the set of reflection and transmission amplitudes obtained from the BdG analysis described in Section~\ref{sec:gs}, from which we compute the transmission eigenvalues $T_n$ and the corresponding shot noise power $S$ and Fano factor, $F$. 
The results reveal distinct trends in the retro (RAR) and specular (SAR) Andreev reflection regimes highlighting how device geometry and interface configuration control noise behavior.

\subsection{GS Interface}
Fig.~\ref{fig:shot_noise1} collects the key trends for the GS interface. Fig.~\ref{fig:shot_noise1a} shows the angle-averaged subgap shot noise, using the definitions of \(r_A^{\mathrm{spec}}(\alpha,\beta)\) and \(r_A^{\mathrm{retro}}(\alpha,\beta)\) (see Appendix \ref{appendix_retrospec}), with \(T=|r_A|^2\) and
\begin{equation}
S(E)\;\propto\;\int_{0}^{\pi/2} T(E,\alpha)\bigl[1-T(E,\alpha)\bigr]\,\mathrm{d}\alpha,
\end{equation}
for specular (blue, solid) and retro (red, dashed) Andreev reflection. It reveals that \(S(E)\) is even in \(E\) because \(\beta(E)=\arccos(E/\Delta)\) enters only through \(\sin\beta\) and \(\cos\beta\). For the specular branch, \(E=0\) implies \(\beta=\pi/2\) and \(\cos\beta=0\), hence \(r_A^{\mathrm{spec}}=0\Rightarrow T=0\) for all \(\alpha\), yielding \(S(0)=0\). As \(\lvert E\rvert\!\to\!\Delta\), \(\beta\!\to\!0\) so \(T(\alpha)\approx\sin^2\alpha\) and the angle integral of \(T(1-T)\) remains finite, producing the nonzero plateau. For the retro branch, \(E=0\) gives \(\sin\beta=1\), so \(T(\alpha)=\sin^2\alpha\) over a broad range of angles, which maximizes the partition factor \(T(1-T)\) and yields a midgap maximum. As \(\lvert E\rvert\!\to\!\Delta\), \(\sin\beta\!\to\!0\), so \(T\!\to\!0\) and the noise vanishes at the gap edges.
Fig.~\ref{fig:shot_noise1b} displays the corresponding Fano factor \(F(E)\). In the RAR regime, (\(E_F\gg E\)), high-transparency Andreev channels dominate and \(F\) is suppressed below its normal-state value. As \(E_F\) is tuned towards the Dirac point, the process crosses over to SAR; the effective transparency is reduced for most angles and \(F\) rises sharply, providing a clear shot noise signature that distinguishes RAR from SAR in a single-interface geometry.

Fig.~\ref{fig:shot_noise_combo} shows angle–resolved shot–noise maps for a GS interface.
In Fig.~\ref{fig:shot_noise_combo:a}, a dark ridge runs along \(\alpha=0\) and at mid-gap \(\beta=\pi/2\) (zero bias),
where the amplitude forces \(T\to 0\) and thus \(S=0\). At intermediate \(\beta\), bright arcs appear near \(|\alpha|\approx \pi/4\), indicating many trajectories with \(0<T<1\). As \(\beta\to 0\) (approaching the gap edge), the angular dependence persists, so a broad set of \(\alpha\) yields finite \(S\).

In Fig.~\ref{fig:shot_noise_combo:b}, zeros occur along \(\alpha=0\) and at \(\beta=0\) (\(|E|\to\Delta\)), where
Andreev reflection is suppressed. Four bright lobes peak near \(\beta=\pm\pi/2\) and \(|\alpha|\approx \pi/4\),
where many modes have \(T\approx 1/2\) and the partition noise approaches its upper bound.

Fig.~\ref{fig:shot_noise_combo:c} shows the contrast
\(\Delta S(\alpha,\beta)=S_{\mathrm{spec}}(\alpha,\beta)-S_{\mathrm{retro}}(\alpha,\beta)\),
which inherits the symmetries \(\Delta S(\alpha,\beta)=\Delta S(-\alpha,\beta)=\Delta S(\alpha,\pi-\beta)\).
Positive regions appear near the gap edges \(\beta\approx 0,\pi\), where \(S_{\mathrm{spec}}\) remains finite while
\(S_{\mathrm{retro}}\to 0\); negative regions concentrate around mid–gap \(\beta\approx \pi/2\) and
\(|\alpha|\approx \pi/4\), where the retro map is largest. Pale contours trace \(\Delta S=0\), i.e., loci where both mechanisms yield equal shot noise.

\begin{figure}[t]
  \captionsetup{justification=raggedright,singlelinecheck=false}
  \centering
  \begin{subfigure}[t]{0.48\linewidth}
    \centering
    \includegraphics[width=\linewidth]{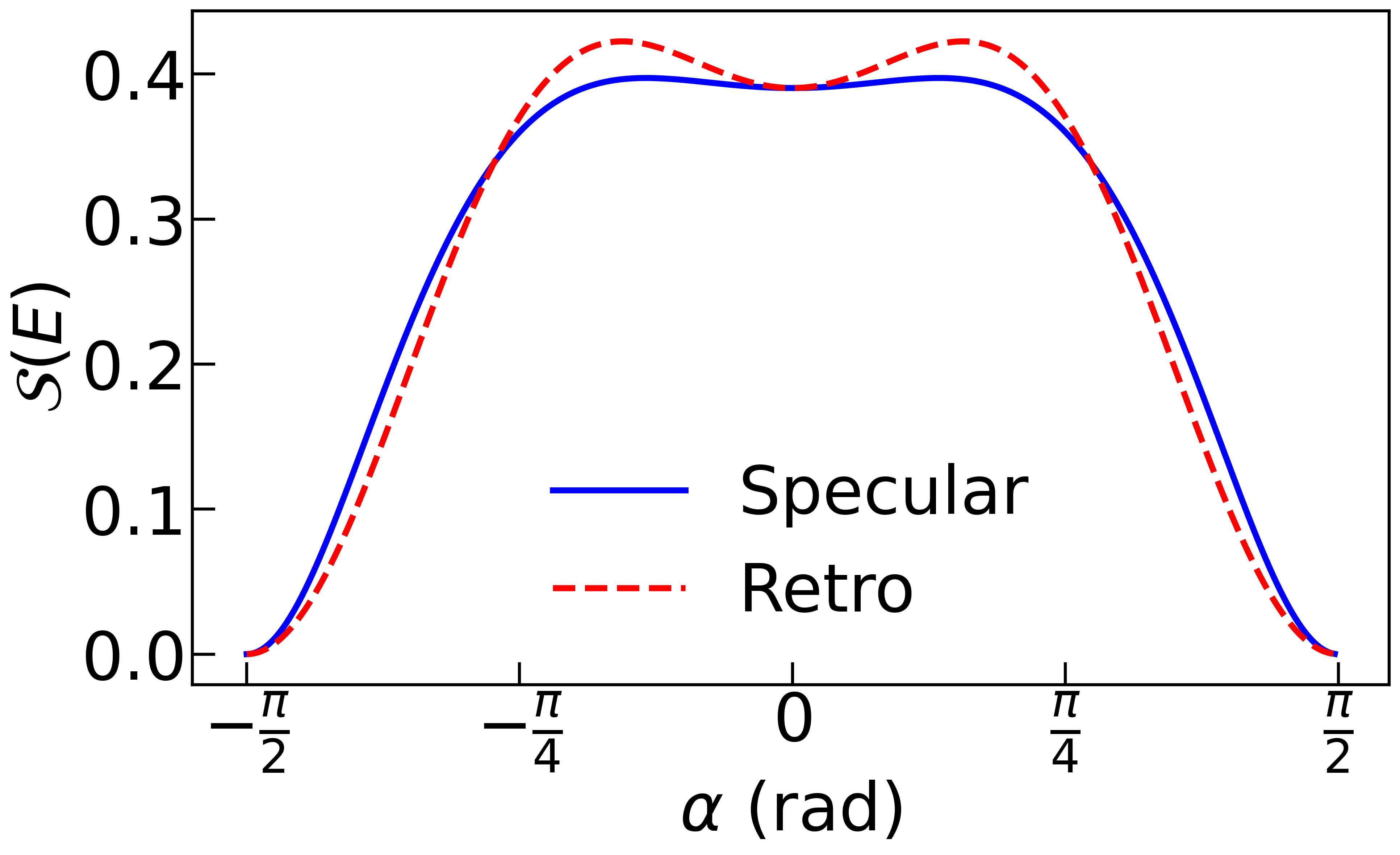}
    \caption{}
    \label{fig:GSG_SN_angle}
  \end{subfigure}\hfill
  \begin{subfigure}[t]{0.48\linewidth}
    \centering
    \includegraphics[width=\linewidth]{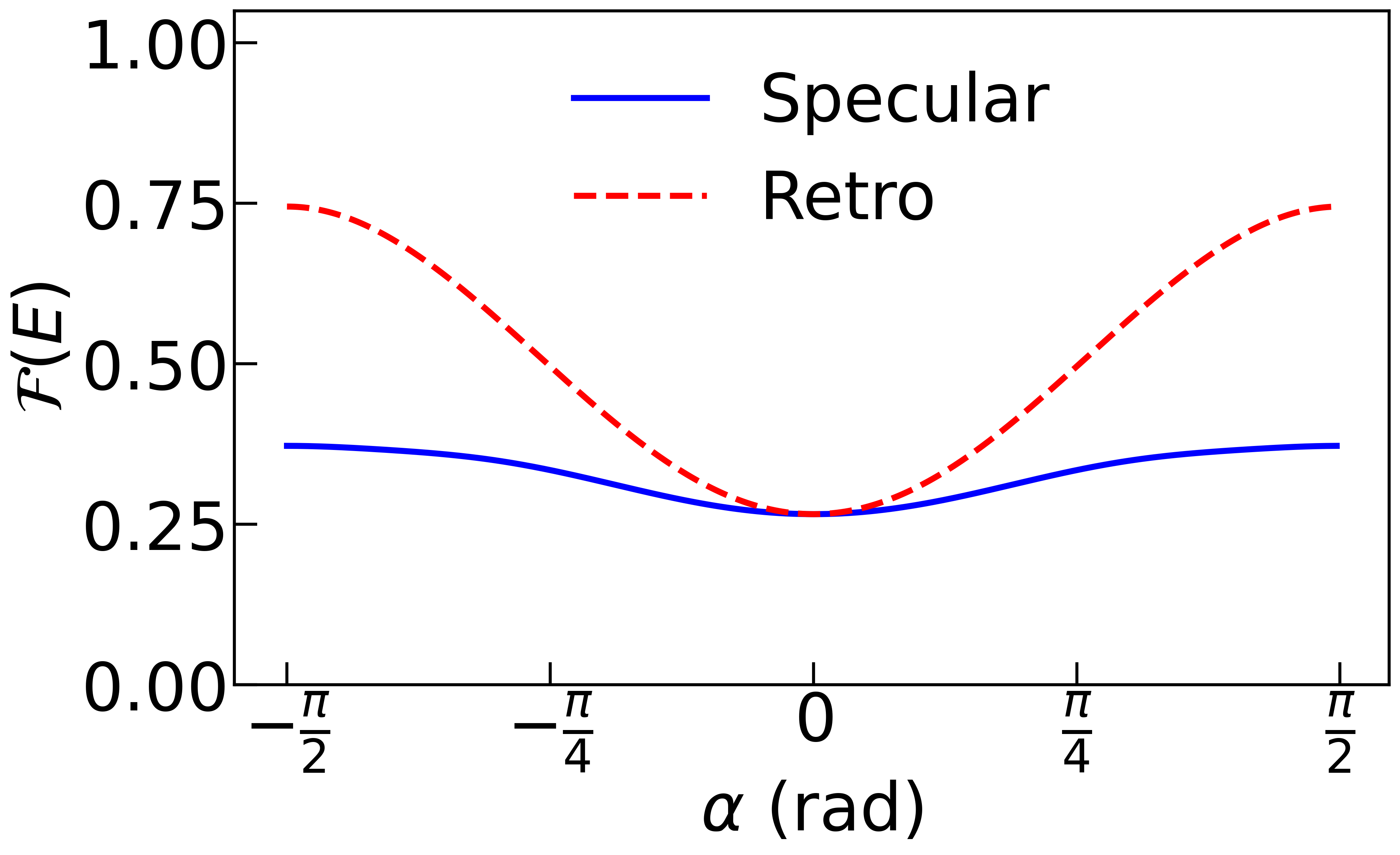}
    \caption{}
    \label{fig:GSG_FF_angle}
  \end{subfigure}
  \caption{(a) Angle-resolved shot noise $S(E)$ and (b) Fano factor $F(E)=S/I$ as a function of incident angle $\alpha$ for specular (blue, solid) and retro (red, dashed) processes in the GSG configuration.}
  \label{fig:GSG_SN_FF}
\end{figure}
\begin{figure}[h]
  \captionsetup{justification=raggedright,singlelinecheck=false}
  \centering
  \begin{subfigure}{0.32\textwidth}
    \centering
    \includegraphics[width=\linewidth]{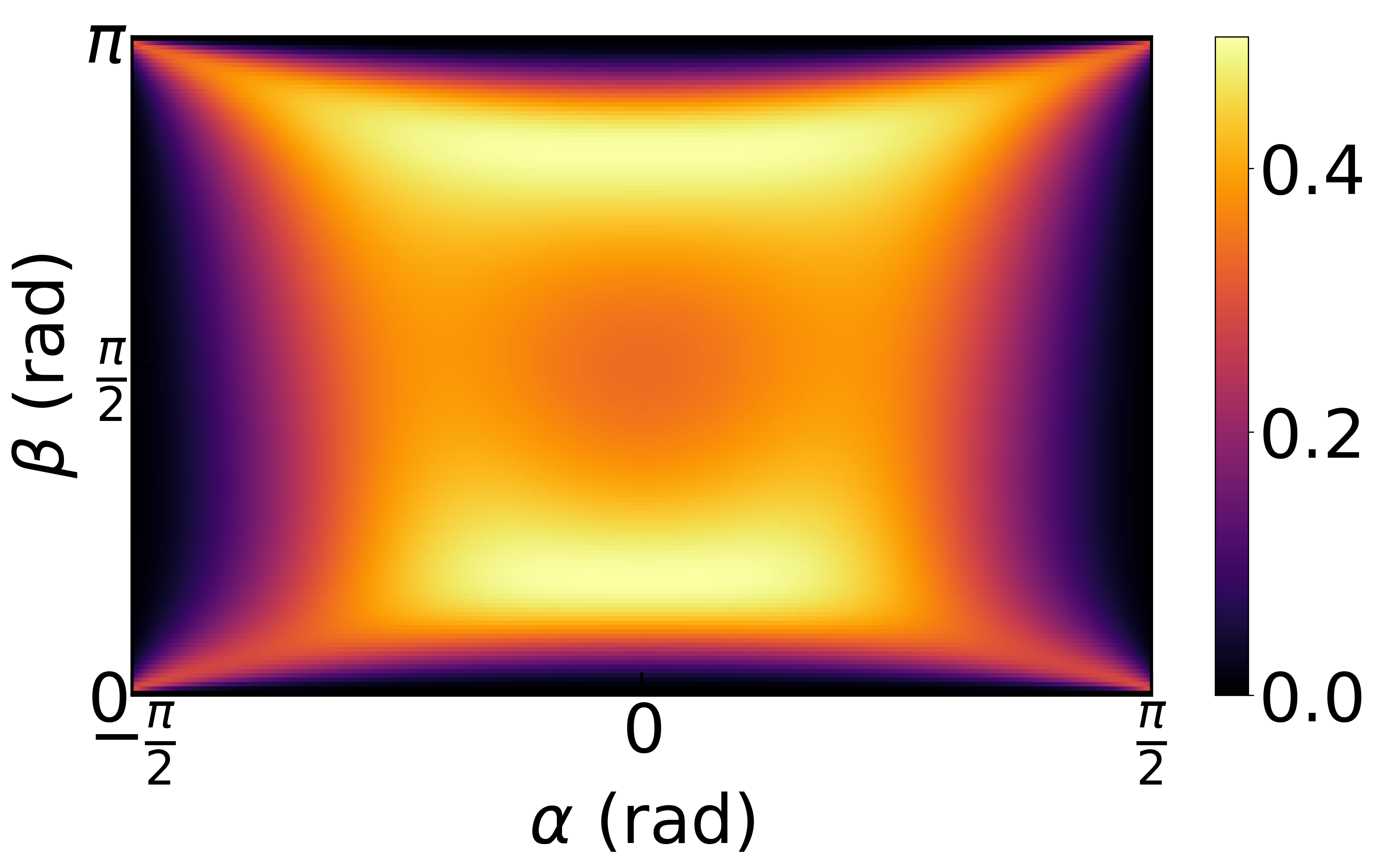}
    \caption{}\label{fig:retro_heatmap}
  \end{subfigure}\hfill
  \begin{subfigure}{0.32\textwidth}
    \centering
    \includegraphics[width=\linewidth]{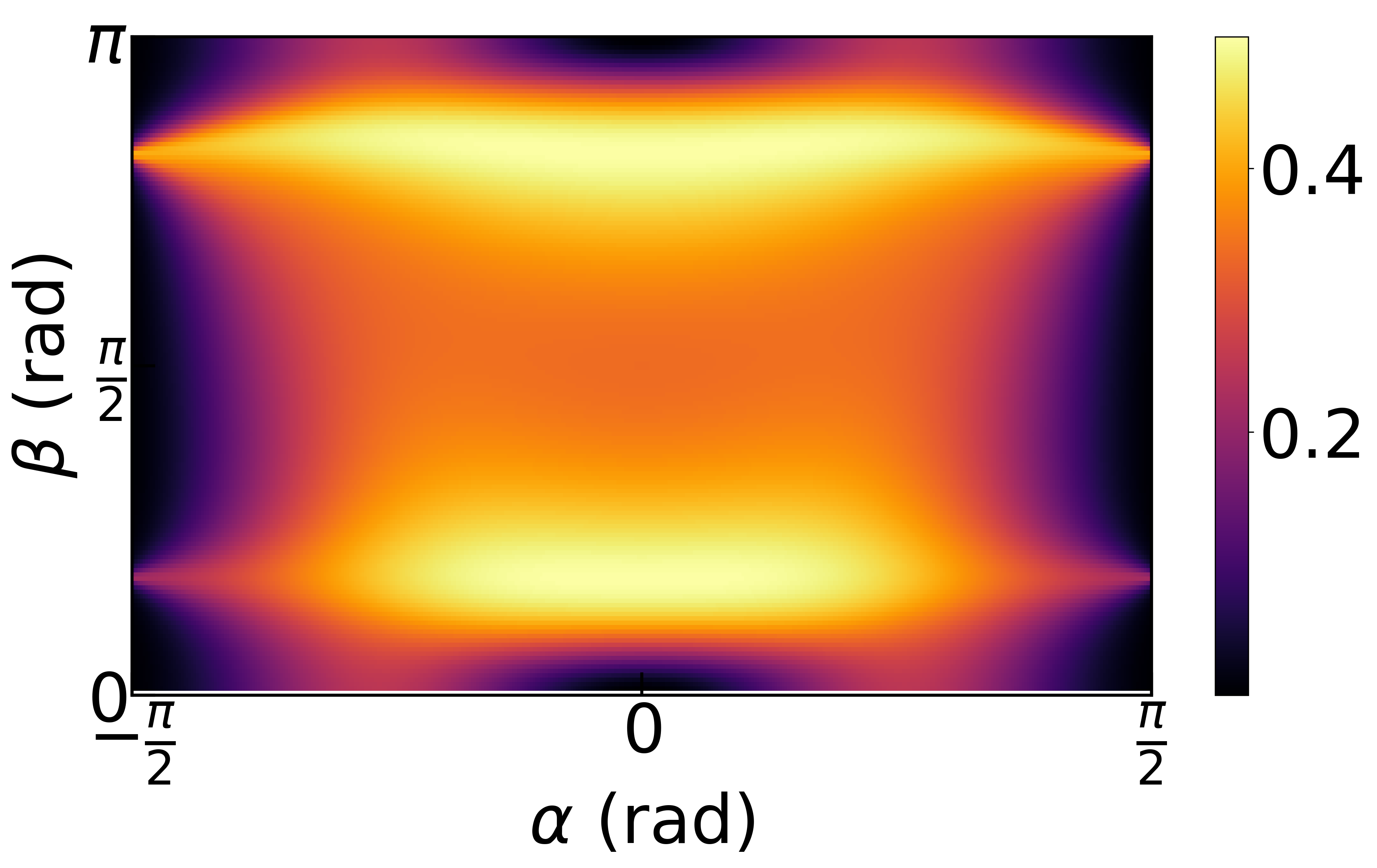}
    \caption{}\label{fig:specular_heatmap}
  \end{subfigure}\hfill
  \begin{subfigure}{0.32\textwidth}
    \centering
    \includegraphics[width=\linewidth]{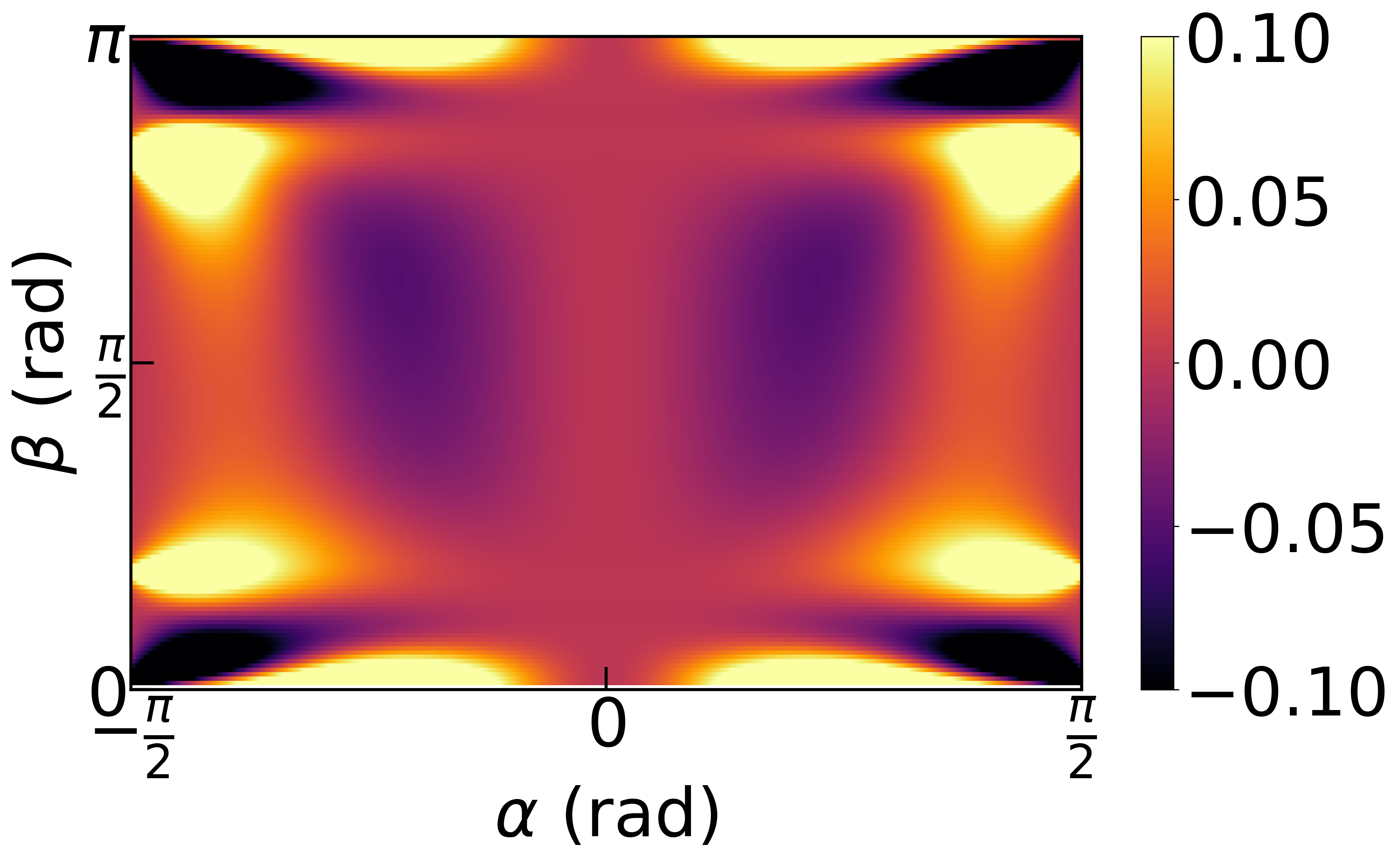}
    \caption{}\label{fig:diff_heatmap}
  \end{subfigure}
  \caption{Angle–resolved shot–noise density \(S(\alpha,\beta)\) for GSG configuration:
  (a) retro $S_{\mathrm{retro}}(\alpha,\beta)$, (b) specular $S_{\mathrm{spec}}(\alpha,\beta)$ and (c) difference \(\Delta S=S_{\mathrm{spec}}-S_{\mathrm{retro}}\).
  Here \(\alpha\in[-\pi/2,\pi/2]\) is the incidence angle and \(\beta=\arccos(E/\Delta)\) encodes the quasiparticle energy. In (a) and (b), warm (cool) colors indicate larger (smaller) noise, while in (c), warm means \(S_{\mathrm{spec}}>S_{\mathrm{retro}}\).}
  \label{fig:three_heatmaps}
\end{figure}
Fig.~\ref{fig:sn_angle_norm_three} shows the normalized angular shot noise $S(\alpha)/S_{\max}(E,T)$ for specular and retro processes at three temperature values, $T=0.2T_c$, $T=0.5T_c$ and $T=0.9T_c$.

At low temperature ($T=0.2T_c$), shown in Fig.~\ref{fig:sn_angle_T02}, the system lies deep in the subgap regime. The retro branch grows rapidly at small incidence angles, reaching its maximum at a relatively low $\alpha$, while the specular branch peaks later and at higher amplitude. The two curves intersect once, marking the transition from retro dominance at small $\alpha$ to specular dominance at larger angles.

At intermediate temperature ($T=0.5T_c$), shown in Fig.~\ref{fig:sn_angle_T05}, the reduced superconducting gap brings the maxima of the two contributions closer together and shifts their crossing toward smaller angles. Both curves exhibit the characteristic form of partition noise: vanishing at normal and grazing incidence, peaking at intermediate $\alpha$, and bounded by the normalization line at unity corresponding to half-open transmission ($T=1/2$). Retro reflection dominates at low-to-moderate angles, while specular reflection overtakes near grazing incidence.

At high temperature ($T=0.9T_c$), shown in Fig.~\ref{fig:sn_angle_T09}, the condition $\Delta(T)<E$ drives $\beta \to 0$, suppressing retro reflection entirely (flat red dashed trace) while the specular (blue, solid) branch persists with a broad arch. This apparent asymmetry reflects applying subgap expressions beyond their range of validity; a proper above-gap Blonder-Tinkham-Klapwijk treatment would make both channels decay together as $\Delta(T)\to 0$. Thus, the "specular-only" behavior at high $T$ is not intrinsic but a modeling artifact, whereas at lower temperatures the expected division - retro dominating small angles, specular dominating large angles - remains consistent with Andreev partitioning across incidence angles.

\begin{figure}[h]
  \centering
  \begin{subfigure}{0.24\textwidth}
    \centering
    \includegraphics[width=\linewidth]{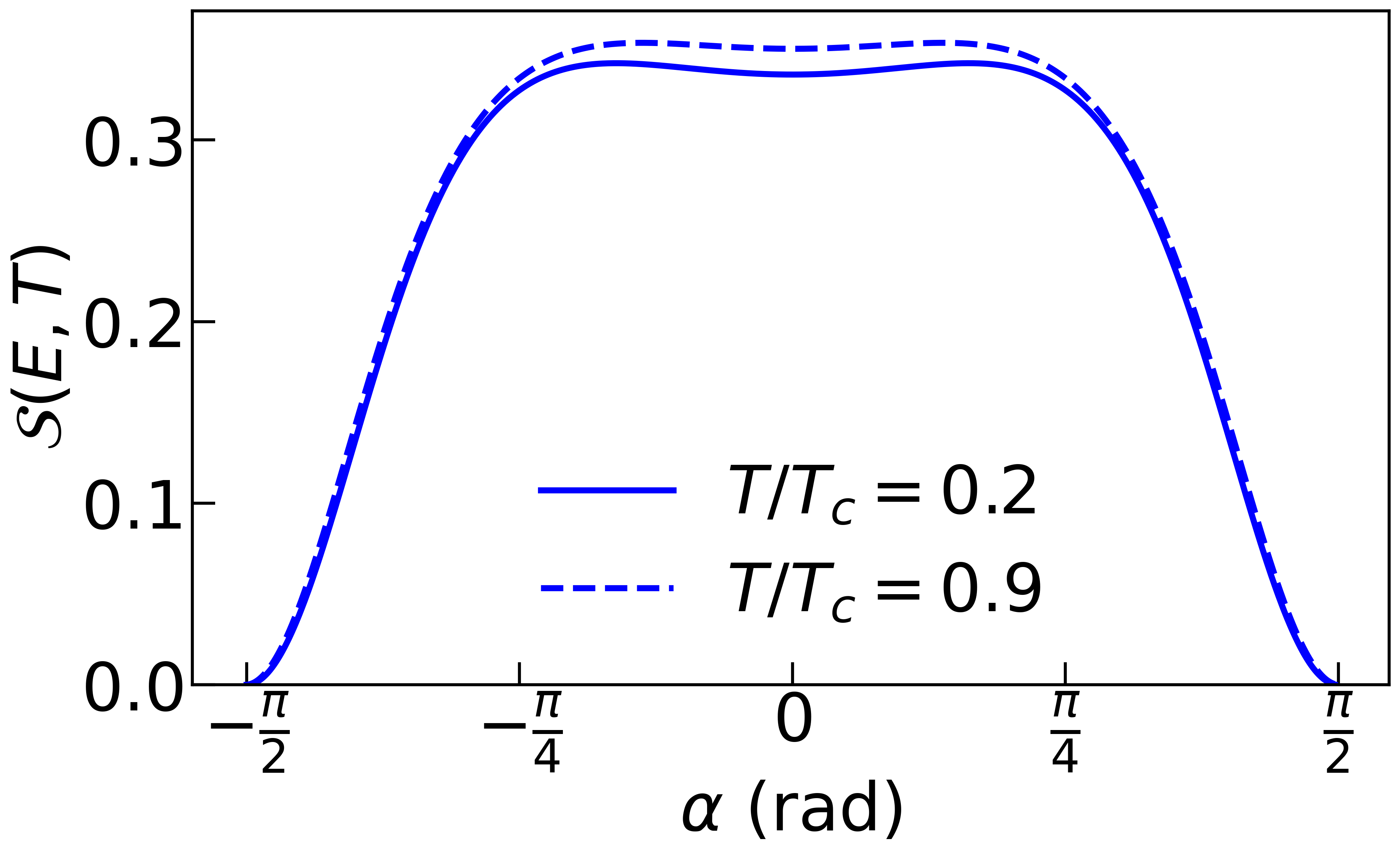}
    \caption{}
    \label{fig:shot-spec}
  \end{subfigure}\hfill
  \begin{subfigure}{0.24\textwidth}
    \centering
    \includegraphics[width=\linewidth]{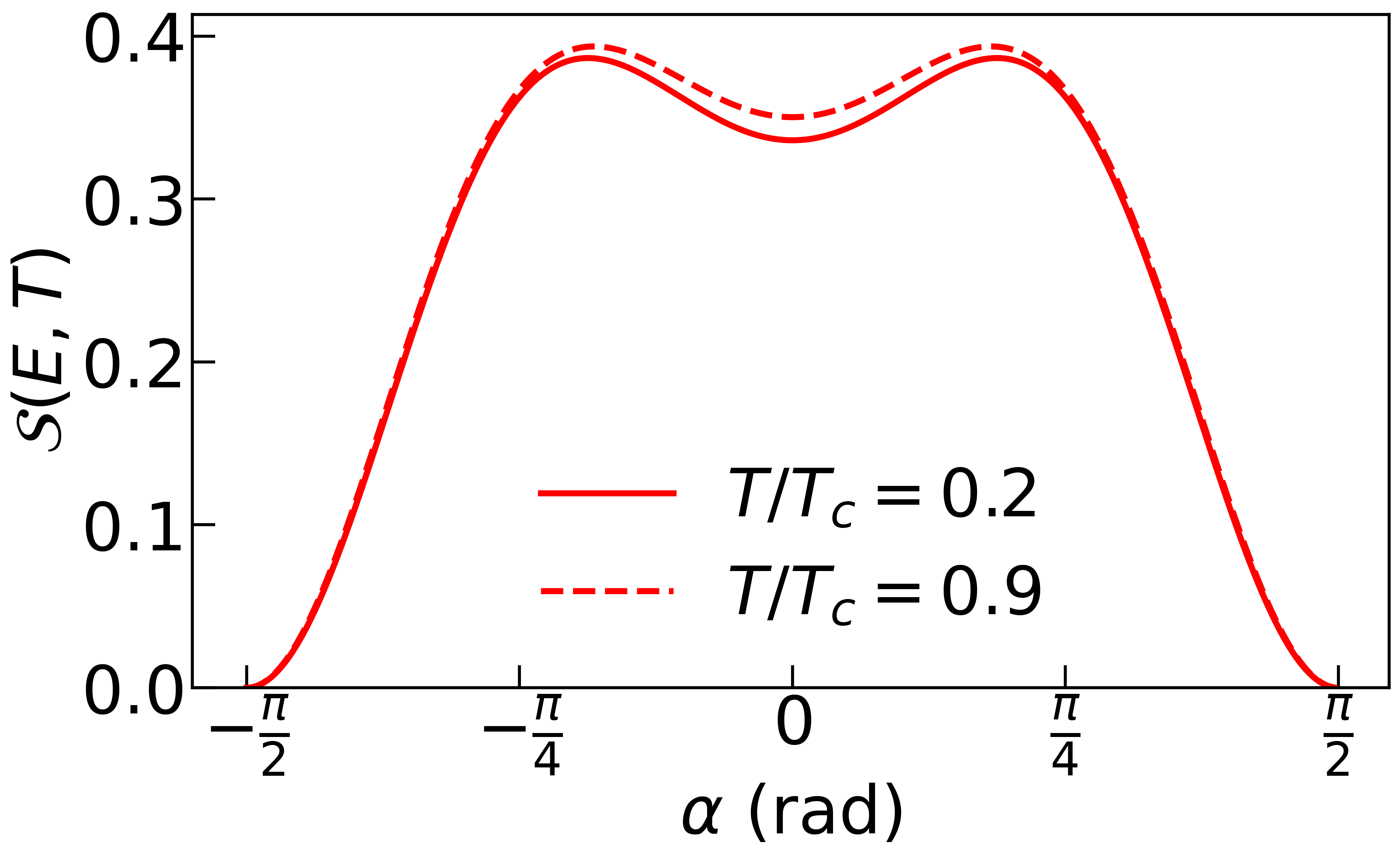}
    \caption{}
    \label{fig:shot-retro}
  \end{subfigure}
  \caption{Angle-resolved shot noise \(S(E,T)\) as a function of
  the incidence angle \(\alpha\) (in radians) for different temperatures for
  (a) specular process and (b) retro process for $T=0.2T_c$ (solid) and $T=0.9T_c$ (dashed).}
  \label{fig:shot-twoT}
\end{figure}

\subsection{GSG Junction}

In the GSG configuration, two GS interfaces separated by a superconducting region of width $d$ give rise to multiple Andreev and normal reflections.
The total transfer matrix $M_{\mathrm{GSG}}$  yields the full set of scattering amplitudes across the device, from which $T_n$ and the Fano factor are computed. 

In Fig.~\ref{fig:GSG_SN_angle}, the angle-resolved shot noise
\(S(E)\) in the GSG cavity is even in \(\alpha\) and collapses near
grazing incidence \(|\alpha|\!\to\!\pi/2\), where transmission becomes either
nearly closed or nearly perfect so that the partition factor \(T(1-T)\)
vanishes. Between these edges both traces rise into a broad, almost flat
plateau with a shallow dip around normal incidence. The retro curve (red,
dashed) sits consistently above the specular one (blue, solid), indicating that retro
trajectories populate a larger set of half-open channels
(\(T\!\approx\!1/2\)) and therefore, generate stronger partition noise. The
specular branch is smoother and slightly lower across most angles, consistent
with its stronger angular filtering inside the cavity.

The corresponding Fano factor \(F(E)=S/I\) in Fig.~\ref{fig:GSG_FF_angle}
mirrors these trends. It is symmetric in \(\alpha\), exhibits a clear minimum
at \(\alpha\!\approx\!0\), and grows toward \(|\alpha|\!\to\!\pi/2\).
Quantitatively, the retro trace climbs to noticeably larger values than the
specular one, reflecting reduced effective transparency for retro paths in the
resonator. Since \(F=\sum_n T_n(1-T_n)\big/\sum_n T_n\), this separation
directly captures the different angular distributions of transmission
eigenvalues generated by the two Andreev mechanisms. This shows that oblique
incidence enhances both noise and Fano factor in the GSG geometry, with retro
AR systematically noisier than specular AR - an angle-resolved signature of how
the two processes partition current within the double-interface cavity.

In Fig.~\ref{fig:retro_heatmap} (retro), a broad bright interior indicates that
many trajectories achieve intermediate transmission (\(T\approx 1/2\)),
producing strong partition shot noise. The intensity weakens towards normal and
grazing incidence and along the gap edges, consistent with the reduction of the
Andreev probability there.

Fig.~\ref{fig:specular_heatmap} (specular) shows a complementary pattern:
pronounced nodes along \(\alpha=0\) and near mid–gap \(\beta\approx\pi/2\),
where the specular Andreev amplitude is suppressed, and bright ridges at oblique
incidence and for \(\beta\) displaced from mid–gap (roughly
\(\beta\approx\pi/4\) and \(3\pi/4\)), reflecting the angular selectivity of
specular conversion.

The contrast is quantified in Fig.~\ref{fig:diff_heatmap}, which plots
\(\Delta S(\alpha,\beta)=S_{\mathrm{spec}}-S_{\mathrm{retro}}\).
Positive lobes appear near the gap edges (\(\beta\approx 0,\pi\)), where the
specular noise remains finite while the retro contribution collapses; negative
regions cluster around mid–gap \(\beta\approx\pi/2\) and
\(|\alpha|\approx\pi/4\), where retro dominates. The pale zero–contours trace
loci where both mechanisms yield comparable noise.

Fig.~\ref{fig:shot-twoT} shows the thermally averaged shot noise
\(S(E,T)\) as a function of the incidence angle \(\alpha\) for two
temperatures, \(T/T_c=0.2\) (solid) and \(T/T_c=0.9\) (dashed) for specular (see Fig.~\subref{fig:shot-spec}) and retro (Fig.~\subref{fig:shot-retro}) AR.
Comparing solid and dashed curves within each panel highlights the temperature
dependence of \(S(E,T)\) across angles: at higher \(T/T_c\)
(dashed), thermal broadening modifies the angular profile relative to the
lower-temperature curves (solid). Separating specular
(Fig.~\ref{fig:shot-twoT}\subref{fig:shot-spec}) and retro
(Fig.~\ref{fig:shot-twoT}\subref{fig:shot-retro}) scattering clarifies how the
two regimes respond differently to temperature over the same range of
incidence angles.

\subsection{SGS Junction}

The SGS Josephson junction replaces the central superconducting region of the GSG device with a graphene segment of length $L$ between two superconducting leads. 
Here, the subgap transport is mediated by Andreev bound states (ABS) whose energies $E(\phi)$ depend on the superconducting phase difference $\phi$ and on $E_F$. 
Using the transfer matrix $M_{\mathrm{SGS}}$, we extract the ABS dispersion and, compute the  shot noise.

\begin{figure}[h]
  \centering
  \begin{subfigure}[t]{0.32\textwidth}
    \centering
    \includegraphics[width=\linewidth]{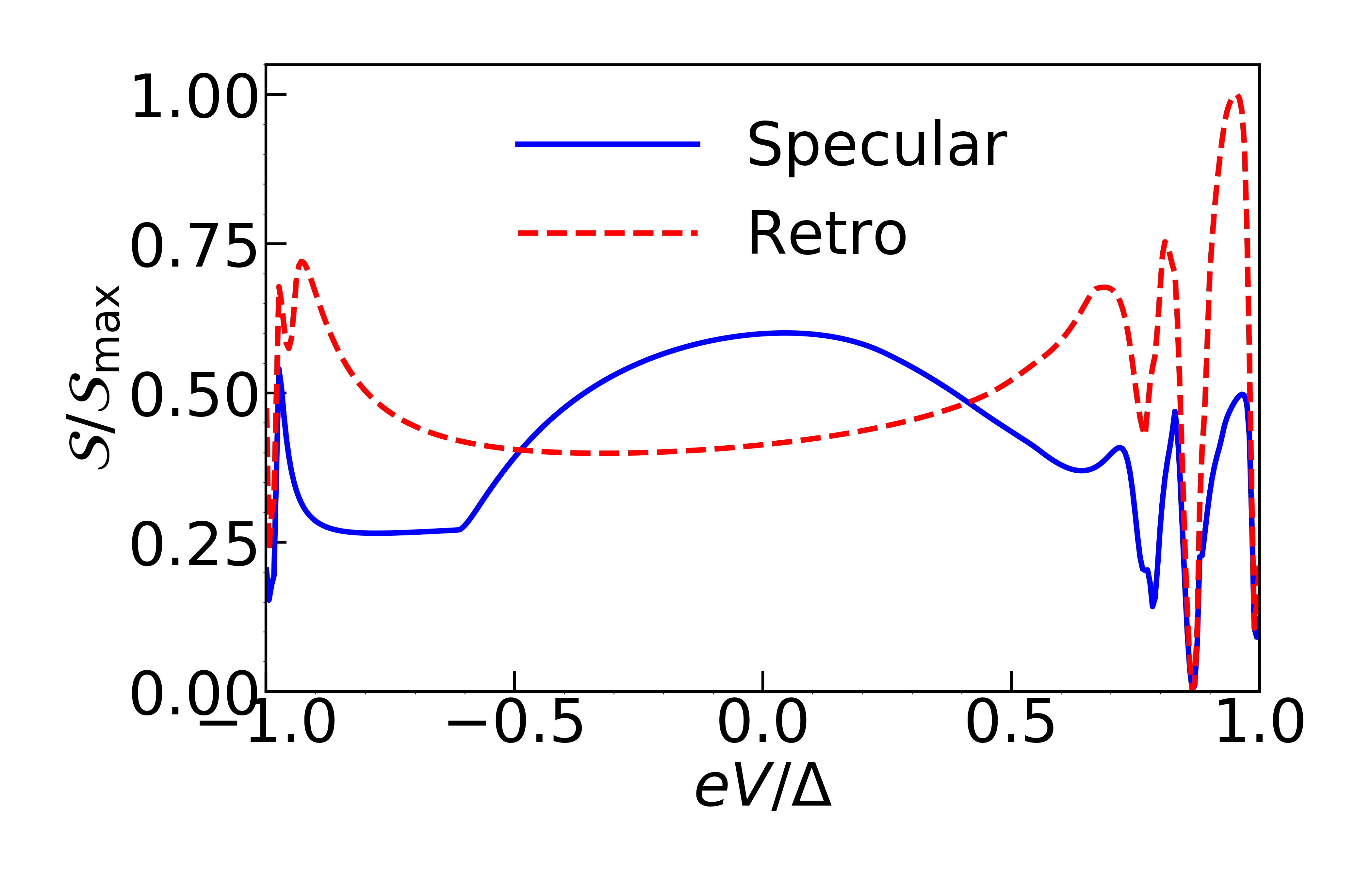}
    \caption{}
    \label{fig:shotnoise}
  \end{subfigure}\hfill
  \begin{subfigure}[t]{0.32\textwidth}
    \centering
    \includegraphics[width=\linewidth]{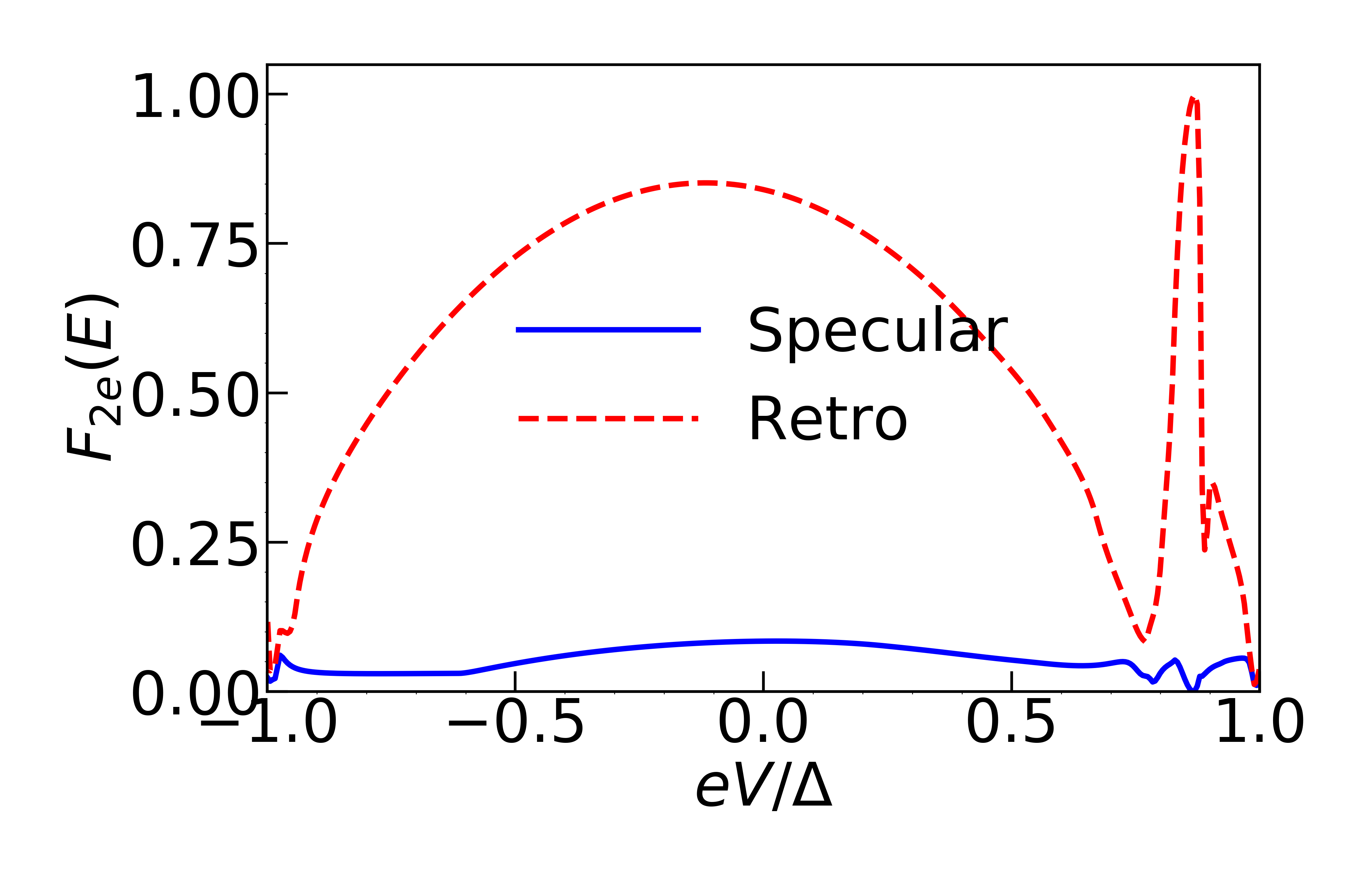}
    \caption{}
    \label{fig:fano2e}
  \end{subfigure}\hfill
  \begin{subfigure}[t]{0.32\textwidth}
    \centering
    \includegraphics[width=\linewidth]{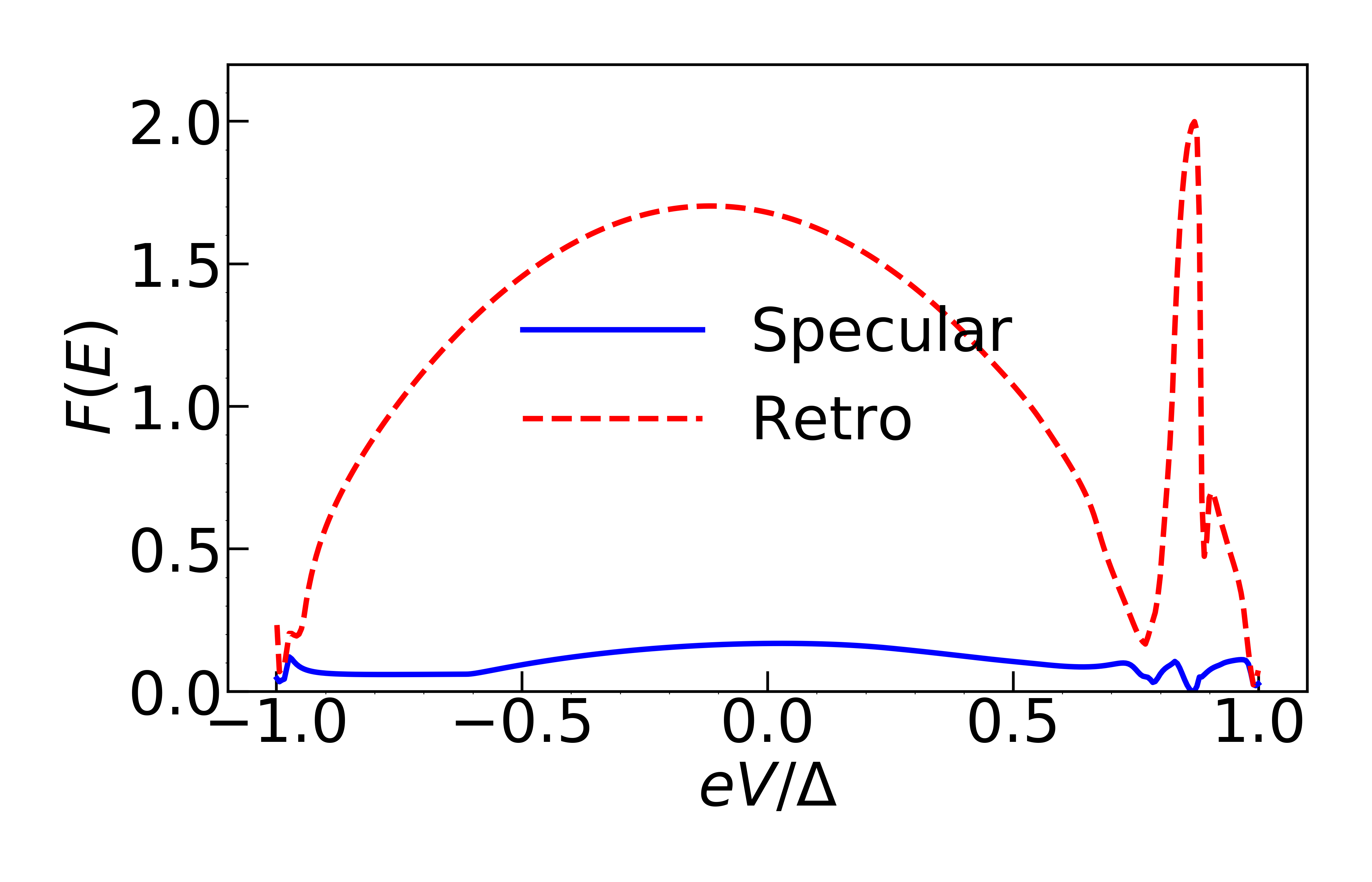}
    \caption{}
    \label{fig:fanoe}
  \end{subfigure}
  \caption{(a) Angle-averaged shot noise, normalized to the global maximum,
    $S/S_{\max}$, (b) Cooper-pair normalization \(F_{2e}(E)=S/(4eI)=F(E)/2\) and (c) Fano factor \(F(E)=S/(2eI)\) as a fucntion of bias \(E=eV\) (in units of \(\Delta\)) for specular (blue, solid) and retro (red, dashed) Andreev branches.}
  \label{fig:fano-shot}
\end{figure}

\begin{figure}[h]
  \centering
  \begin{subfigure}[t]{0.32\textwidth}
    \centering
    \includegraphics[width=\linewidth]{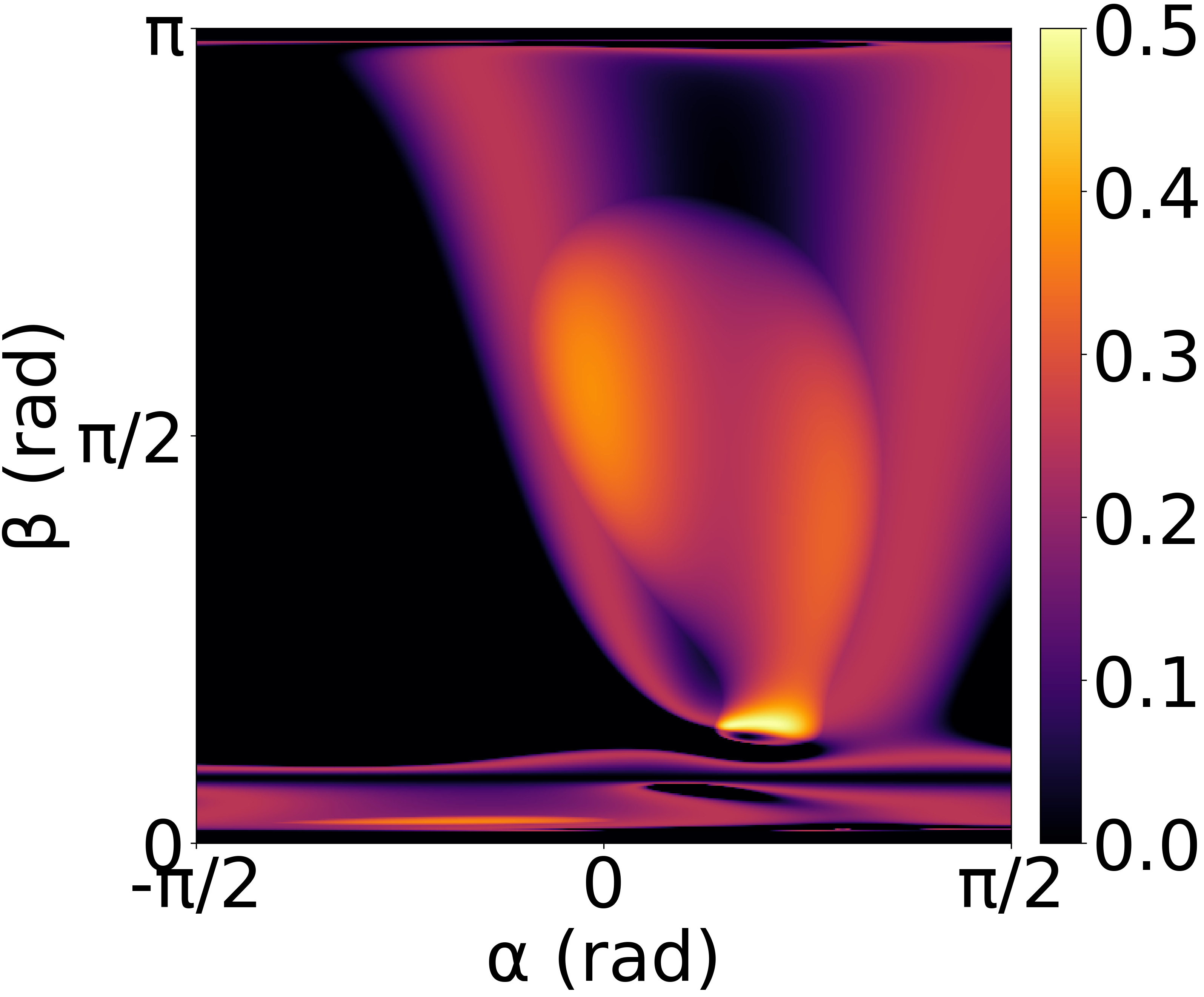}
    \caption{}
    \label{fig:spec}
  \end{subfigure}\hfill
  \begin{subfigure}[t]{0.32\textwidth}
    \centering
    \includegraphics[width=\linewidth]{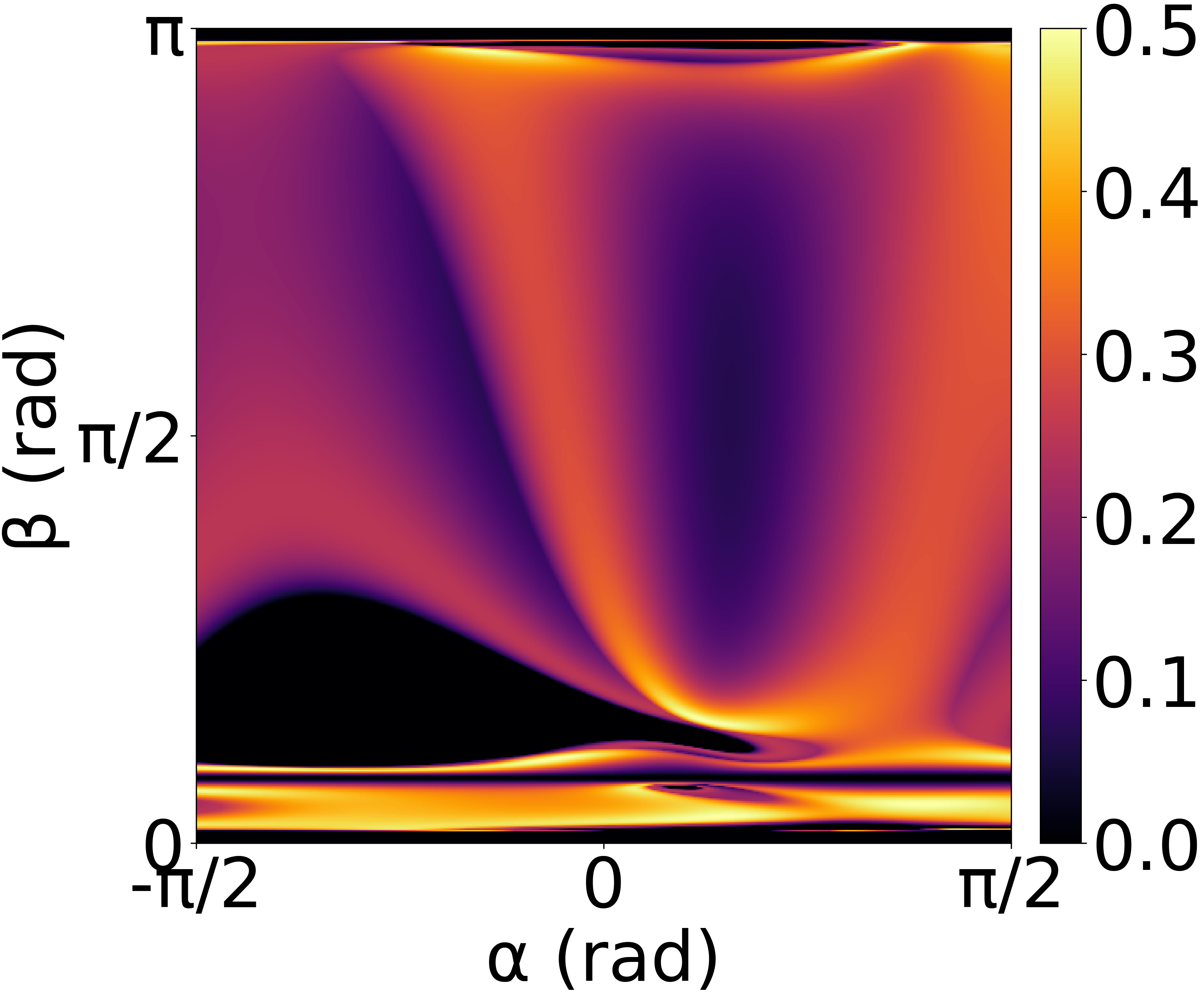}
    \caption{}
    \label{fig:retro}
  \end{subfigure}\hfill
  \begin{subfigure}[t]{0.32\textwidth}
    \centering
    \includegraphics[width=\linewidth]{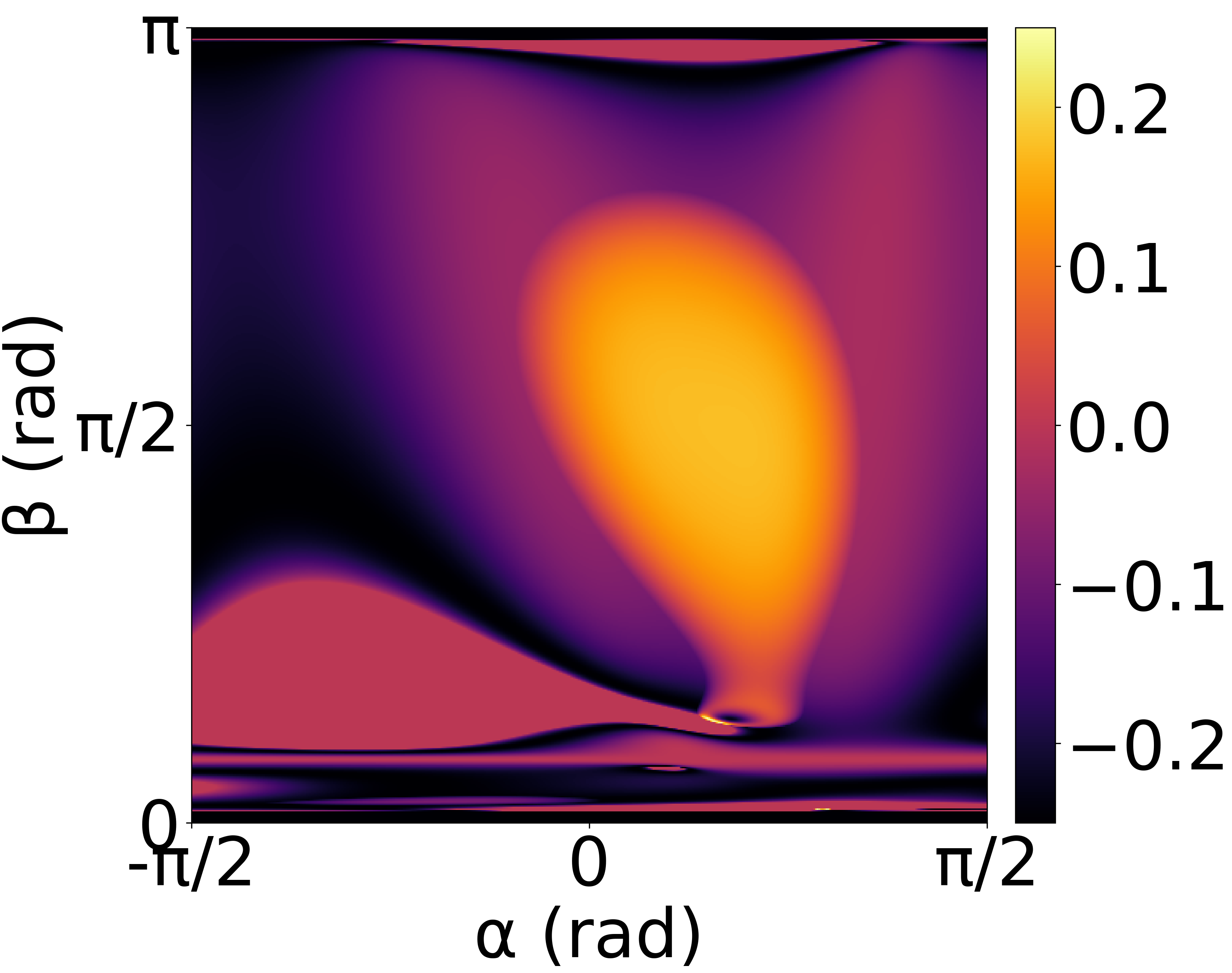}
    \caption{}
    \label{fig:diff}
  \end{subfigure}

  \caption{Shot-noise maps  \(S(\alpha,\beta)\) as a function of incidence angle \(\alpha\) (horizontal) and superconducting phase angle \(\beta\) (vertical) for (a) specular process, (b) retro process and (c) difference \(S_{\mathrm{spec}}-S_{\mathrm{retro}}\). The same color scale is used for Figs.~\ref{fig:spec} and \ref{fig:retro}; Fig.~\ref{fig:diff} uses a symmetric scale around zero.}
  \label{fig:shotnoise_panels}
\end{figure}

Fig.~\ref{fig:fano-shot} summarizes the energy dependence of the current–noise
characteristics. For reference,
\[
S(E)=2e\,I(E)\,F(E), \qquad \text{with } F(E)\ \text{(Fano factor)},
\]
with \(F(E)\) set by energy– and angle–dependent reflection and transmission amplitudes.

In Fig.~\ref{fig:fano-shot}\,\subref{fig:shotnoise}, both branches are finite due to
Andreev processes. The specular curve exhibits a pronounced dome around zero
bias, reflecting stronger energy sensitivity of the effective Andreev probability
when the reflected hole occupies the opposite band. The retro curve is flatter
across much of the subgap window, consistent with hole retracing and more nearly
deterministic scattering; see also the corresponding Fano factors in
Figs.~\ref{fig:fano-shot}\,\subref{fig:fano2e}–\subref{fig:fanoe}.

Both traces in Fig.~\ref{fig:fano-shot}\,\subref{fig:shotnoise} develop sharp
structure - dips, spikes, and oscillations - arising from rapid variations of the
superconducting coherence factors and phase-accumulation/interference as channels
open at the gap. These edge features carry over to
Fig.~\ref{fig:fano-shot}\,\subref{fig:fanoe}, where \(F(E)\) shows corresponding
peaks/dips set by the same coherence physics.

Fig.~\ref{fig:fano-shot}\,\subref{fig:fano2e} displays \(F_{2e}(E)=S/(4eI)=F/2\),
the Cooper-pair normalization appropriate when Andreev transport dominates; values
approaching unity indicate nearly perfect \(2e\) transfer. In
Fig.~\ref{fig:fano-shot}\,\subref{fig:fanoe}, the electron-unit normalization
\(F(E)=S/(2eI)\) can exceed unity in Andreev-dominated windows,
consistent with the effective charge \(e^\ast\!\approx\!2e\). Across both panels,
the specular branch tends to yield larger low-bias partition noise, whereas the
retro branch is comparatively flatter in the subgap region and sharper near the
gap edge. The contrasting trends in shot noise and Fano factor provide an experimentally accessible fingerprint to distinguish specular from retro Andreev processes in these junctions.

In Fig.~\ref{fig:spec},
a bright lobe appears at moderate \(\beta\) near normal incidence
\(\alpha \approx 0\), signaling strong partitioning between normal and Andreev
channels 
Distinct nodes are visible along \(\alpha=0\) and near mid-gap
values of \(\beta\), where the specular Andreev amplitude is suppressed. At more
oblique angles, the intensity increases, forming broad ridges that reflect the
angular selectivity of specular conversion.

In Fig.~\ref{fig:retro}, 
 the intensity around \(\alpha\approx0\) is
reduced and a wider dark region emerges where \(S\approx 0\), consistent with
either weak reflection or nearly deterministic scattering. Bright lobes develop
at intermediate \(|\alpha|\), indicative of many trajectories with
\(T\approx 1/2\) and thus, strong partition noise, while the signal weakens
towards grazing incidence and near the gap edges.

In Fig.~\ref{fig:diff}, 
the difference \(\Delta S(\alpha,\beta)=S_{\mathrm{spec}}-S_{\mathrm{retro}}\)
highlights where each mechanism dominates: warm (positive) regions cluster near
the gap edges, where the specular noise remains finite while the retro
contribution collapses; cool (negative) regions concentrate around mid-gap and
\(|\alpha|\!\sim\!\pi/4\), where retro is largest. The pale zero contours trace
the loci where both mechanisms yield comparable shot noise.

This shows that specular AR dominates the fluctuation landscape near normal incidence and intermediate \(\beta\), whereas retroreflection is comparatively quieter except along thin, resonant corridors at oblique angles. This complementary structure provides a
clear, angle– and energy–resolved fingerprint for distinguishing specular from retro Andreev processes.

\begin{figure}[h]
  \centering
  \begin{subfigure}[t]{0.32\textwidth}
    \centering
    \includegraphics[width=\linewidth]{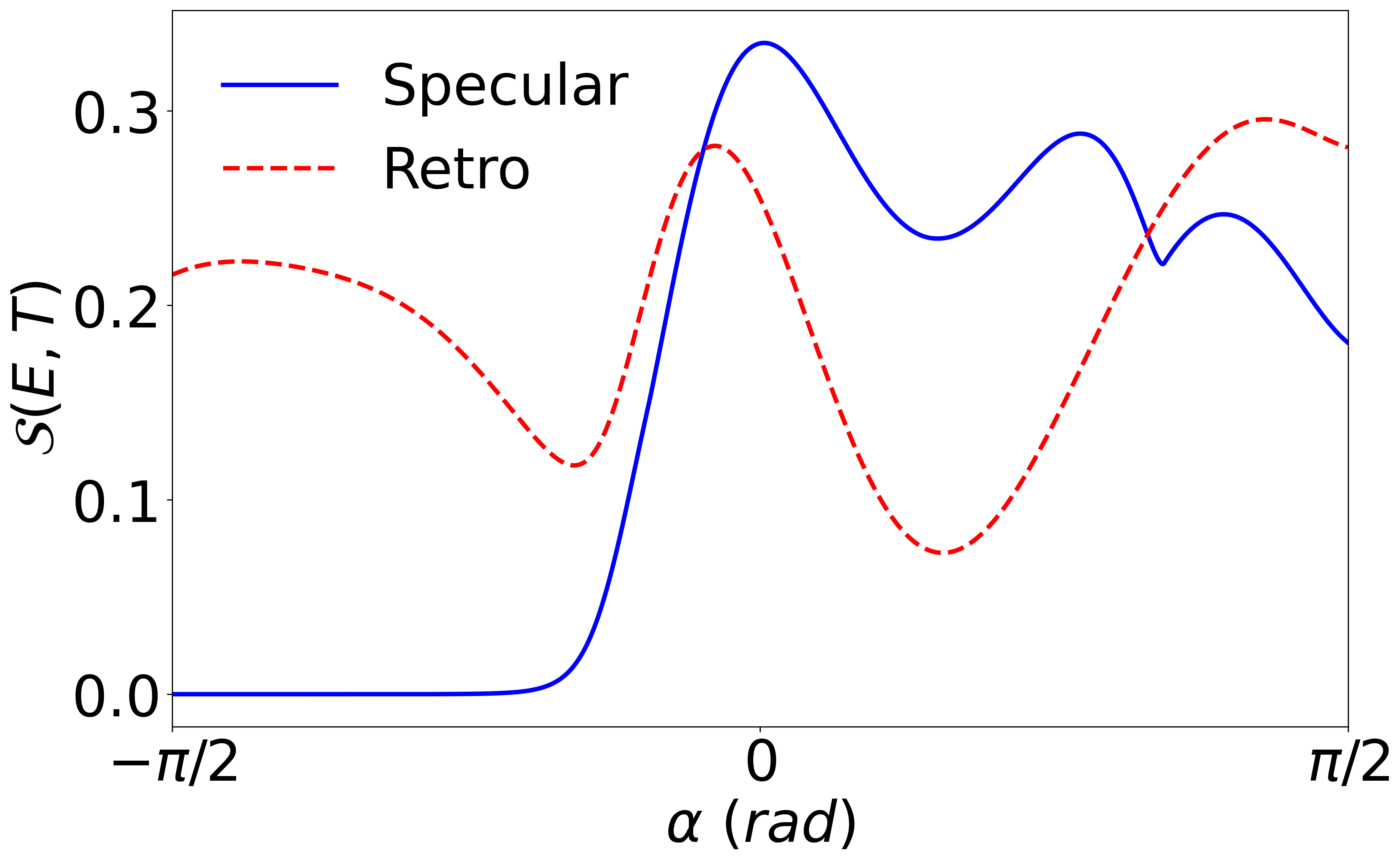}
    \caption{}
    \label{fig:shot-alpha-02}
  \end{subfigure}\hfill
  \begin{subfigure}[t]{0.32\textwidth}
    \centering
    \includegraphics[width=\linewidth]{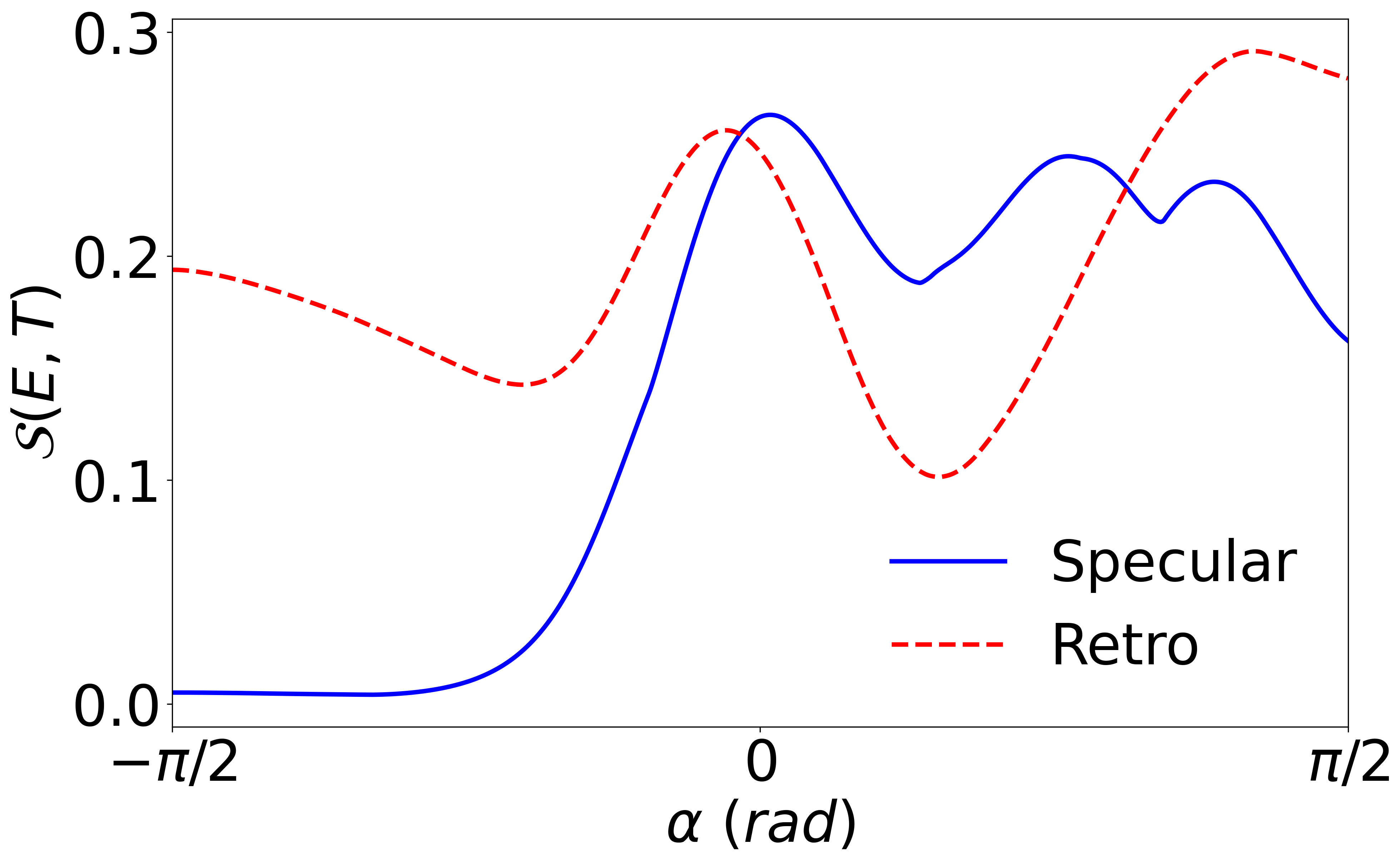}
    \caption{}
    \label{fig:shot-alpha-05}
  \end{subfigure}\hfill
  \begin{subfigure}[t]{0.32\textwidth}
    \centering
    \includegraphics[width=\linewidth]{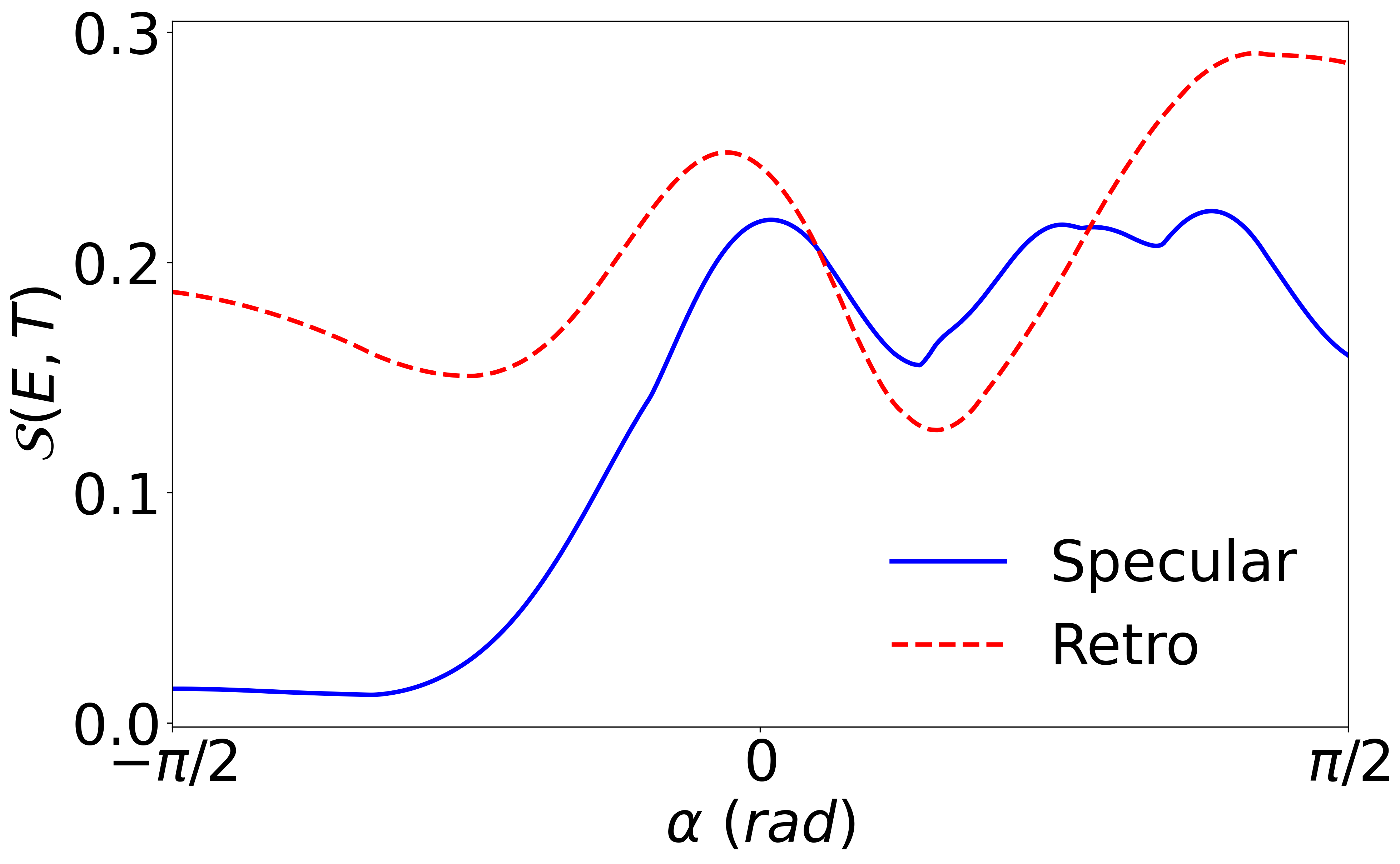}
    \caption{}
    \label{fig:shot-alpha-09}
  \end{subfigure}
  \caption{Angle dependence of the (dimensionless) shot noise $S(E,T)$ as a function of incidence angle $\alpha$ at three reduced temperatures (a) $T/T_c=0.2$, (b) $T/T_c=0.5$ and (c) $T/T_c=0.9$ for specular Andreev channel (blue, solid) and retro Andreev channel (red, dashed).}
  \label{fig:shot-alpha-temp}
\end{figure}

Fig.~\ref{fig:shot-alpha-temp} summarizes the angular shot-noise profiles for specular (blue, solid) and retro (red, dashed) Andreev processes. It
uses the transmission probability $\mathcal{T}$ and the Fermi function $f$. The gap follows the mean-field law
$\Delta(T)=\Delta_0\sqrt{1-(T/T_c)^2}$, so increasing $T$ both reduces $\Delta$ and broadens the thermal kernel $f(1-f)$, which together smear angular interference and lower the overall magnitude of $S(E,T)$.

At low temperature (Fig.~\ref{fig:shot-alpha-02}, $T/T_c=0.2$), the specular contribution dominates near normal incidence ($\alpha\!\approx\!0$), showing a pronounced peak and mild oscillations. The retro branch is comparatively suppressed around $\alpha\!=\!0$ and contributes more strongly at oblique angles.

At intermediate temperature (Fig.~\ref{fig:shot-alpha-05}, $T/T_c=0.5$), the specular peak is reduced and the oscillations are smooth, while the retro curve gains relative weight at larger $|\alpha|$. The crossing between the two traces shifts towards larger $\alpha$, indicating a progressive rebalancing between specular and retro channels.

Near the critical temperature (Fig.~\ref{fig:shot-alpha-09}, $T/T_c=0.9$), both curves flatten and approach each other over most angles, consistent with the diminishing superconducting gap. In the limit $T\!\to\!T_c$, one expects $S(E,T)$ to approach its normal-state form, set by $\mathcal{T}\,[1-\mathcal{T}]$ alone, with minimal angular contrast.

Overall, the trends are governed by the shrinking $\Delta(T)$ (which shifts $\beta$ and the Andreev amplitudes) and by thermal averaging: lower $T$ enhances coherent angular features and favors specular Andreev conversion near normal incidence; higher $T$ suppresses these features and brings the specular and retro contributions closer together.

\section{Conclusions and Outlook}\label{sec:conclusion}

In summary, we have shown that shot noise provides a clear and experimentally accessible differentiator between specular and retro Andreev reflection in graphene–superconductor heterojunctions. Within a unified BdG scattering framework, the angle- and energy-resolved noise maps $S(\alpha,\beta)$ and the associated Fano factor exhibit robust, qualitative contrasts: in the specular regime, $S$ features nodal lines at normal incidence and near mid-gap, with bright off-center ridges at oblique angles; in the retro regime, $S$ remains broadly elevated across the interior and is suppressed mainly near geometric and gap edges. These complementary patterns are stable against variations in interface transparency and gating, and are crisply captured by the difference maps $\Delta S(\alpha,\beta)=S_{\mathrm{spec}}-S_{\mathrm{retro}}$.

The predicted noise fingerprints: suppressed Fano factor in the retro-dominated regime, enhanced and gate-tunable Fano factor in the specular regime, and distinct temperature/angle trends, are readily accessible to experimental verification with existing techniques. High-mobility, hBN-encapsulated monolayer graphene with one-dimensional edge contacts to Al/Pb/NbN (or MoRe/NbN for larger gaps) routinely yields clean GS/GSG/SGS junctions. Cross-correlated, low-frequency (hundreds of kHz–MHz) cryogenic noise spectroscopy performed in the subgap window $k_B T \ll eV \ll \Delta(T)$ enables accurate extraction of $F=S_I/(2eI)$ (or $F_{2e}=S_I/(4eI)$). A single gate sweep across the Dirac point at fixed subgap bias should reveal the RAR$\to$SAR crossover: $F$ is minimized deep in the doped (RAR-dominated) regime and rises toward charge neutrality (SAR-dominated), while controlled temperature sweeps drive the two behaviors to merge as $\Delta(T)$ decreases. Practical pitfalls like charge-puddle disorder, a finite interface barrier $Z$, and electron heating, can be mitigated via hBN encapsulation, low-resistance edge contacts, careful RF filtering and thermalization, and in-situ calibration against the field-driven normal state. Taken together, these considerations place our proposal well within present experimental reach and motivate targeted noise spectroscopy in graphene–superconductor hybrids and their multi-terminal extensions.

\appendix

\section{Wavefunction Solutions and Matching}\label{appendix_wavefuncmatch}

To obtain the reflection amplitudes at the graphene–superconductor interface, we construct explicit wavefunction solutions in both regions and impose boundary conditions at $x = 0$ \cite{BTK1982,Beenakker1992NSreview,TinkhamBook1996,deGennes1999}.

In the normal graphene region ($x > 0$), the eigenstates of the Dirac Hamiltonian in the electron and hole sectors correspond to plane-wave solutions with well-defined transverse momentum $k_y$ (set by the conserved parallel wavevector) and incidence angles $\alpha$ (for electrons) and $\alpha'$ (for holes) \cite{CastroNeto2009,Beenakker2008GrapheneColloquium}. For an electron-like quasiparticle incident from the right ($-$) or left ($+$), the four-component Nambu spinor is
\begin{equation}
\psi_e^{\pm} =
\frac{e^{\pm i k_x x + i k_y y}}{\sqrt{2\cos\alpha}}
\begin{pmatrix}
e^{\mp i\alpha/2} \\
\pm e^{\pm i\alpha/2} \\
0 \\
0
\end{pmatrix},
\end{equation}
while a hole-like excitation propagating to the right ($+$) is described by
\begin{equation}
\psi_h^{+} =
\frac{e^{i k_x x + i k_y y}}{\sqrt{2\cos\alpha'}}
\begin{pmatrix}
0 \\
0 \\
e^{-i\alpha'/2} \\
-e^{i\alpha'/2}
\end{pmatrix}.
\end{equation}
Here, $\alpha$ and $\alpha'$ are related to the longitudinal and transverse momenta through $\sin\alpha = k_y / k_e$ and $\sin\alpha' = k_y / k_h$, where $k_e$ and $k_h$ are the electron and hole wavevectors in graphene.

In the superconducting region ($x < 0$), subgap quasiparticles are evanescent and decay away from the interface. The corresponding modes are written as
\begin{equation}
\psi_s^{\pm} =
A  e^{\pm \kappa x + i k_y y}
\begin{pmatrix}
e^{\mp i\beta} \\
\pm e^{\pm i(\gamma - \beta)} \\
e^{\mp i\phi} \\
\pm e^{\pm i(\gamma - \phi)}
\end{pmatrix},
\end{equation}
where $\kappa$ is the decay constant in the superconducting region, $\beta$ is the superconducting coherence angle defined via $\cos\beta = E/\Delta_0$, $\phi$ is the macroscopic superconducting phase, and $\gamma$ is the angle of propagation inside the superconductor \cite{TinkhamBook1996,deGennes1999}. The upper (lower) two components of each spinor correspond to the electron (hole) sector.

The boundary condition at the GS interface requires the continuity of the wavefunction at $x = 0$. For an incident electron from the graphene side, this condition reads
\begin{equation}
\psi_e^{-} + r_e , \psi_e^{+} + r_h , \psi_h^{+}
= a , \psi_s^{-} + b , \psi_s^{+}
\end{equation}
where $r_e$ and $r_h$ are the normal and Andreev reflection amplitudes, and $a$ and $b$ are coefficients for the decaying modes in the superconductor \cite{BTK1982,Zaitsev1984,KupriyanovLukichev1988}.

Solving the matching equations yields \cite{Beenakker2006SpecAR,Nilsson2007,Linder2008,BlackSchaffer2008}
\begin{align}
r_h &= \frac{\sqrt{\cos\alpha , \cos\alpha'} , e^{-i\phi}}
{\cos\beta \cos\left(\frac{\alpha' - \alpha}{2}\right)
+ i \sin\beta \cos\left(\frac{\alpha' + \alpha}{2}\right)}, \\
r_e &= i , \frac{\sin\beta \sin\left(\frac{\alpha' - \alpha}{2}\right)
- \cos\beta \sin\left(\frac{\alpha' + \alpha}{2}\right)}
{\cos\beta \cos\left(\frac{\alpha' - \alpha}{2}\right)
+ i \sin\beta \cos\left(\frac{\alpha' + \alpha}{2}\right)}.
\end{align}

\section{Retro and Specular Reflection}\label{appendix_retrospec}

The above expressions describe both retro Andreev reflection (RAR) and specular Andreev reflection (SAR), depending on the relative magnitudes of the Fermi energy $E_F$ and the excitation energy $E$.

\paragraph*{Retro reflection ($E_F \gg E$).} In this intraband case, the hole resides in the same band as the incident electron, and $\alpha' \approx -\alpha$ \cite{BTK1982,Beenakker1992NSreview}:
\begin{align}
r_h &= \frac{\cos\alpha , e^{-i\phi}}
{\cos\beta \cos\alpha + i \sin\beta}, \
r_e &= \frac{-i \sin\beta \sin\alpha}
{\cos\beta \cos\alpha + i \sin\beta}.
\end{align}

\paragraph*{Specular reflection ($E_F \ll E$).} In this interband process, the reflected hole lies in the opposite band, and $\alpha' \approx \alpha$ \cite{Beenakker2006SpecAR,CastroNeto2009,Beenakker2008GrapheneColloquium}:
\begin{align}
r_h &= \frac{\cos\alpha , e^{-i\phi}}
{\cos\beta + i \sin\beta \cos\alpha}, \
r_e &= \frac{-\cos\beta \sin\alpha}
{\cos\beta + i \sin\beta \cos\alpha}.
\end{align}
These compact forms make explicit the dependence of the reflection amplitudes on the quasiparticle incidence angle, excitation energy, Fermi level, and superconducting phase.

\bibliographystyle{apsrev4-2} 
\bibliography{ref} 

\end{document}